\documentclass{aa}  

\usepackage{graphicx}
\usepackage{txfonts}
\usepackage{soul}
\usepackage{xcolor}
\newcommand{\del}[1]{\setstcolor{red}\st{#1}}

\begin{document}

\title{Estimating masses of supermassive black holes in
  active galactic nuclei from the H$\alpha$ emission line}
\titlerunning{AGN black hole masses from the H$\alpha$ emission line}

   \author{E.\ Dalla\ Bont\`{a}
          \inst{1,}\inst{2,}\inst{3}
          \and
          B.M.\ Peterson\inst{4}
          \and
          C.J.\ Grier\inst{5}
          \and
          M.\ Berton\inst{6}
          \and          
          W.N.\ Brandt \inst{7,}\inst{8,}\inst{9}
          \and
          S. Ciroi\inst{1,}\inst{2}
          \and
          E.M.\ Corsini\inst{1,}\inst{2}
          \and
          B.\ Dalla Barba\inst{10,}\inst{11}
          \and
          R. Davies\inst{12}
          \and 
          M.\ Dehghanian\inst{13}
          \and
          R.\ Edelson\inst{14}
          \and
          L. Foschini\inst{11}
          \and
          D.\ Gasparri\inst{15}
          \and 
           L.C.\ Ho\inst{16,}\inst{17}
          \and
          K.\ Horne\inst{18}
          \and
          E.\ Iodice\inst{19}
          \and
          L.\ Morelli\inst{15}
          \and
          A.\ Pizzella\inst{1,}\inst{2}
          \and
          E.\ Portaluri\inst{20}
          \and
          Y.\ Shen\inst{21,}\inst{22}
          \and
          D.P.\ Schneider\inst{7,}\inst{8}
          \and
          M.\ Vestergaard\inst{23,}\inst{24}
}
          
   \institute{Dipartimento di Fisica e Astronomia ``G. Galilei,''
     Univerisit\`{a} di Padova, Italy \\ 
   \email{elena.dallabonta@unipd.it}  
     \and
   INAF -- Osservartorio Astronomico di Padova, Vicolo dell'Osservatorio 5,    I-35122 Padova, Italy  
    \and
    Jeremiah Horrocks Institute, University of Central Lancashire, Preston, PR1 2HE, UK 
   \and 
   Retired 
   \and  
   Department of Astronomy, 
   University of Wisconsin -- Madison,
   Sterling Hall,    475 N.\ Charter St., 
   Madison, WI 53706-1507, USA  
   \and
  European Southern Observatory, Alonso de Córdova 3107, 
  Casilla 19, Santiago 19001, Chile 
   \and
   Department of Astronomy and Astrophysics,
   Eberly College of Science, The
   Pennsylvania State University, 525 Davey
   Laboratory, University Park, PA 16802, USA  
   \and
   Institute for Gravitation and the Cosmos,
   The Pennsylvania State University,
   University Park, PA 16802, USA 
   \and
   Department of Physics, 
   The Pennsylvania State University,
   525 Davey Laboratory, University Park, PA 16802, USA 
   \and
   Università degli studi dell’Insubria, Via Valleggio 11, Como 22100, Italy 
   \and 
   Osservatorio Astronomico di Brera, Istituto Nazionale di Astrofisica (INAF), Via E. Bianchi 46, Merate, 23807, Italy 
   \and
      Max Planck Institute for Extraterrestrial Physics, Giessenbachstrasse, 85741 Garching bei München, Germany 
   \and
   Department of Physics, Virginia Tech, Blacksburg, VA 24061, USA 
   \and
   Eureka Scientific Inc., 2452 Delmer St. Suite 100,
   Oakland, CA 94602, USA 
   \and
   Instituto de Astronomía y Ciencias Planetarias, Universidad de Atacama, Copayapu 485, Copiapó, Chile 
   \and
    Kavli Institute for Astronomy and Astrophysics, Peking University,
    Beijing 100871, China 
    \and
     Department of Astronomy, School of Physics, Peking University,
     Beijing 100871, China  
    \and
    SUPA School of Physics and Astronomy, 
    North Haugh, St. Andrews, KY16 9SS, Scotland, UK 
   \and
   INAF – Osservatorio Astronomico di Capodimonte, Via Moiariello 16, 80131 Napoli, Italy 
   \and
   INAF -- Osservatorio Astronomico d'Abruzzo, via M. Maggini snc, I-64100, Teramo, Italy 
   \and
   Department of Astronomy, University of Illinois Urbana-Champaign, Urbana, IL 61801, USA 
   \and
   National Center for Supercomputing Applications, University of Illinois Urbana-Champaign, Urbana, IL 61801, USA 
   \and
   DARK, Niels Bohr Institute, University of Copenhagen,
   Jagtvej 155, DK-2200 Copenhagen, Denmark 
   \and
   Steward Observatory, University of Arizona, 933 N.\ Cherry Ave., Tucson, AZ 85721, USA  
   } 
\authorrunning{E.\ Dalla Bont\`{a} et al.}
   
   \date{Received 25 October 2024; accepted 23 January 2025}

 
  \abstract
 {}
   {The goal of this project is to construct 
 an estimator for the masses of
 supermassive black holes in active galactic nuclei (AGNs)
 based on the broad H$\alpha$ emission line.}
  {We make use of published reverberation mapping
  data. We remeasure all H$\alpha$ time
  lags from the original data as we find
  that often the reverberation measurements are improved
  by detrending the light curves.}
   {We produce mass estimators that require
   only the H$\alpha$ luminosity 
   and the width of the H$\alpha$ emission line
   as characterized by either the FWHM or
   the line dispersion.}
   {It is possible, on the basis of a single
   spectrum covering the H$\alpha$ emission line,
   to estimate the mass of the central
   supermassive black hole in AGNs, taking into account
   all three parameters believed to affect
   mass measurement --- luminosity, line width,
   and Eddington ratio. The typical formal accuracy in
 such estimates is of order $0.2$ -- $0.3\,{\rm dex}$
 relative to the reverberation-based masses.
 }


   \maketitle
%

   \section{Introduction}

Astrophysical masses are measured by observations of how they accelerate nearby
objects. In the case of supermassive black holes at
the centers of massive galaxies, masses are measured by modeling the
dynamics of stars \citep[e.g.,][]{vanderMarel98, Cretton99, Gebhardt03,
Thomas04, Valluri04, Sharma14}, gas 
\citep[e.g.,][]{Macchetto97, Bower98, Davies04a, Davies04b, deFrancesco08, Hicks08,EDB09, Davis13} 
or megamasers \citep[e.g.,][]{Wagner13, vdB16, Kuo20}
on spatially resolved
scales. In the case of 
some relatively nearby active galactic nuclei (AGNs), 
the broad-line emitting gas can be spatially resolved
with interferometry \citep{GRAVITY18, GRAVITY20,
GRAVITY21a, GRAVITY21b, GRAVITY24}.  In other cases,
the motions of
gas on spatially unresolved scales can be modeled for mass measurement
by the process of reverberation mapping \citep{Pancoast11, Grier13b,
Pancoast14, Grier17a}. The
ultraviolet, optical, and infrared spectra of AGNs are dominated by
the presence of strong, Doppler-broadened emission lines that vary in
flux in response to continuum variations that 
arise on accretion-disk
scales. By mapping the response of the line-emitting gas as a function
of line-of-sight velocity and time delay relative to the continuum
variations, the kinematics of the line-emitting region and the mass of
the central black hole can be determined.  
However, the technical demands of  velocity-resolved (i.e., `` two-dimensional'')
reverberation mapping (RM) are formidable compared to 
simpler measurement of the mean emission-line response time, or time lag, for an entire emission line $\tau$ and the
emission-line width $\Delta V$, i.e., ``one-dimensional RM''
\citep{Blandford82, Peterson93, Peterson14};
compared to one-dimensional RM, two-dimensional RM 
requires more accurate relative flux calibration
(including flat-fielding) as well as more accurate
wavelength calibration and consistent spectral
resolution. It has therefore been more common to
measure the one-dimensional response of the emission-line
and the line width and combine
these to determine the black-hole mass
\begin{equation}
\label{eq:virial}
M_{\rm BH} = f \left( \frac{\Delta V^2\,c\tau}{G}\right) =f\mu,
\end{equation}
where the quantity in parentheses has units of mass and is known as
the ``virial product'' $\mu$ and is based on the two observables,
line width and mean time delay. Under most circumstances
(e.g., except when the continuum radiation or 
emission-line response is highly asymmetric),
the mean time delay translates immediately into the mean radius of the
line-emitting region, $R = c\tau$. Parameters that are not directly
measured by this method, such as the inclination of the line-emitting
region, are subsumed into the dimensionless factor $f$. In the absence
of knowledge of these other parameters, it is common to use a mean
value $\langle f \rangle$ based on other statistical estimates of the
masses, nearly always the $M_{\rm BH}$--$\sigma_*$
relationship  between black-hole mass and bulge stellar velocity
dispersion.  This relationship was first recognized
in quiescent galaxies
\citep{Ferrarese00,Gebhardt00a} but has also
been identified in active galaxies
\citep{Gebhardt00b,Ferrarese01,Nelson04,Watson08,
Grier13a}.

Even one-dimensional RM is resource-intensive, typically
requiring a sequence of at least some 30 -- 50 high-quality
spectroscopic observations over a suitable span of time 
(typically several times the light-crossing time $\tau=R/c$)
with an appropriate sampling rate
(sampling interval typically around $0.5R/c$ or less)
and source variations that are conducive to 
successful reverberation detection. 
Fortunately, however, RM has shown that the
emission-line region radii inferred from lags correlate with many
different luminosity measures $L$ approximately as $L \propto
R^{1/2}$, thus enabling estimates of the central black-hole mass from
a single spectroscopic observation. As the RM database has grown over
time, it has become clear that this ``radius--luminosity relation'' or $R$--$L$ relation is
oversimplified and there is at least one more parameter that affects
the radius of the line-emitting region (hereafter referred to as the
``broad-line region'' or BLR). The additional parameter is generally
thought to be Eddington ratio
\citep[e.g.,][]{Du16, Grier17b, Du18, Du19, Martinez19, Alvarez20},
i.e., the ratio of true
accretion rate to the Eddington accretion rate.  There is a long
history of using the $R$--$L$ scaling relation to estimate the BLR
radius from a measured luminosity and combining this with the
emission-line width to estimate the mass via equation
(\ref{eq:virial}), much of which we reviewed in our earlier paper
\citep[][hereafter Paper I]{EDB20}. Our
investigation reported in Paper I supports the 
conclusion that Eddington ratio is the missing parameter in
the $R$--$L$ relationship and demonstrates that this can be
effectively be taken into 
account. In Paper I, we focused on updating
the $R$--$L$ relations for H$\beta$ and \ion{C}{IV}\,$\lambda1549$; the
former because it has by far the best established RM database, and the
latter because it affords a probe of the higher-redshift universe, and
has been, we think unfairly as we discuss in Paper I, 
deemed by some authors to be
insufficiently reliable for mass estimates. 

In the present work, we
focus on the other strong emission line in the optical,
H$\alpha\,\lambda6563$.
Compared to other strong broad emission lines in AGN spectra,
H$\alpha$ has been relatively neglected in RM studies. There are
several reasons for this:
\begin{enumerate}
\item The sensitivity of the UV/optical detectors generally employed in
  ground-based RM studies limits the redshift range over which
  H$\alpha$ can be observed. In the samples discussed in this paper,
  the highest redshift AGNs are at redshifts $z \la 0.15$.
\item The low space density of local highly luminous AGNs combined
  with cosmic downsizing means that the luminosity range that can be
  studied via H$\alpha$ reverberation is limited compared to other
  broad emission lines. In the samples
  discussed here, there is only one AGN (3C\,273 = PG\,1226+023) with
  rest-frame 5100\,\AA\ luminosity at 
  $L(5100\,{\textrm \AA}) =\lambda L_{\lambda}(5100\,{\rm \AA}) > 10^{45}\,\mbox{erg\,s$^{-1}$}$ and a bare handful
  with $L(5100\,{\textrm \AA}) > 10^{44}\,\mbox{erg\,s$^{-1}$}$.
\end{enumerate}
Other deficiencies are relative to H$\beta$ (in some cases, but not
all, H$\beta$ and H$\alpha$ are observed simultaneously):
\begin{enumerate}
\item The amplitude of emission-line flux variability is generally higher in H$\beta$   than in H$\alpha$ \citep[e.g.,][]{Peterson04},
which makes the variations easier to detect and characterize.
\item The continuum underneath H$\alpha$\ has more host-galaxy
  starlight contamination than that under H$\beta$, so the continuum
  variations are apparently stronger in the H$\beta$ region of the
  spectrum and the starlight correction to the continuum luminosity at
  H$\alpha$\ is much larger and thus uncertainties are 
  more impactful.
\item In many, but not all, cases, the highest fidelity relative flux
  calibration in the H$\beta$ spectral region is achieved by assuming that
  the [\ion{O}{III}]\,$\lambda\lambda 4959, 5007$ fluxes are constant
  on reverberation timescales.  These lines are more clearly separated
  from H$\beta$ than potential narrow-line calibration sources around
  H$\alpha$ (specifically, [\ion{N}{II}]\,$\lambda\lambda 6548, 6583$
  or [\ion{S}{II}]\,$\lambda\lambda 6716, 6731$). The [\ion{N}{II}]
  lines in particular are much harder to separate from the
  H$\alpha$\ broad emission, which compromises them as internal flux
  calibrators and complicates measuring the broad H$\alpha$ line width
  accurately. For two-dimensional reverberation studies (i.e., those that enable constructions of a 
  velocity-delay map), the [\ion{N}{II}] lines can be
  especially problematic.
\item At some modest redshifts, the H$\alpha$ profile is badly
  contaminated by atmospheric absorption bands (i.e., A-band and
  B-band), and accounting for this is not trivial.
\end{enumerate}
However, recent developments in the study of nearby AGNs at
high-angular resolution in the near-IR with both ground-based (e.g.,
GRAVITY at the VLTI) and space-based (JWST) telescopes has led to a
renewed interest in reverberation results for H$\alpha$ for direct
comparison with mass determinations based on angularly resolved
methods.  For this reason, we decided to reconsider the issue of
estimating AGN black hole masses based on the H$\alpha$ emission
line. Our methodology largely follows that of Paper I. For consistency
with Paper I, we assume $H_0 = 72$\,km\,s$^{-1}$\,Mpc$^{-1}$,
$\Omega_{\rm matter} =0.3$, and $\Omega_{\Lambda} = 0.7$.

\section{Observational database and methodology}
\subsection{Data}
As in Paper I, we employ two high-quality databases for this investigation
\begin{enumerate}
\item We have collected spectra, line-width measurements, and time series
  for reverberation-mapped AGNs that have appeared in the literature up
  through 2019. The objects included here are those from Paper I that also have H$\alpha$
  results available.
\item We have included sources from the SDSS Reverberation Mapping
  Project \citep[][hereafter SDSS-RM, or more compactly, SDSS]{Shen15}.
  While Paper I included only results from the first year of the project,
  here we examined the six-year database described by
  \cite{Shen24}, though as we explain below, only the first two years of spectroscopic monitoring plus a previous year of
  photometric monitoring
  are relevant to the present investigation.
\end{enumerate}
Whenever possible, we use line-width measurements and flux or luminosity measurements from the published sources. In some cases
where we had ready access to the data (notably the
Palomar--Green quasars from \citealt{Kaspi00}), we measured
the line widths ourselves.
In all cases, however, we
remeasured the emission-line lags using the interpolated 
cross-correlation methodology \citep{Gaskell87} as implemented by
\cite{Peterson98} and modified by \cite{Peterson04}. We chose to remeasure all the SDSS lags for two reasons:
\begin{enumerate}
\item As described by \cite{Edelson24}, it is important to examine 
the effects of ``detrending'' the light curves. Detrending means either fitting a low-order polynomial to the light curve and subtracting it from the data or convolving the light curve with a broad function, such as a Gaussian: either will remove the
longest-term trends from the data. We do this because
variations on timescales much longer than reverberation timescales
can, because they contain so much power, lead to overestimates of the reverberation response time scale. Here we attempt a simple linear detrending, following \cite{Edelson24}, of the line and continuum light curves and use the time-series that gives the ``best'' results (generally defined by the smallest uncertainties in the lags).
Typically we find that shorter light curves are unaffected by
detrending, but in longer light curves the effects can be important.
\item In the case of SDSS data, we restricted our analysis to the first two years of spectroscopic observations 
($56660 <  \mathrm{MJD} < 57195$)
plus a preceding 
single year ($56358 < \mathrm{MJD} < 56508$)
of continuum measurements. Because the SDSS quasars with H$\alpha$ reverberation measurements are all
fairly local and low luminosity, the sparse sampling of the 
continuum at earlier epochs and the continuum and
lines at later epochs only adds noise to the 
cross-correlation results.
\end{enumerate}

The data drawn from the literature are presented in Table \ref{table:RMDBHa}. Additional parameters associated with each source
are drawn from Table~A1 of Paper I 
\footnote{Associations between sources in Table \ref{table:RMDBHa}
and Table A1 of Paper I are
obvious, except in the case of Mrk\,6. The three data sets
used here were from MJDs 49250--49872, 49980--50777, and 
53611--54803.}.
As noted above, all lags were remeasured,
but luminosities, adjusted to our adopted cosmology, and line widths are taken from the published sources.
Some line-width measures have been flagged
by the original investigators as being particularly
uncertain, usually because of various data quality
issues. These values are denoted by preceding colons and
are not used in any of the statistical analysis.

Table \ref{table:SDSSHa} presents the parameter values for
the SDSS sample. Some additional necessary parameters appear in
Table A2 of Paper I. Luminosities are based on parameters
given by \citet{Shen24}, line widths are from \citet{Wang19}, and  the H$\alpha$ rest-frame
lags are based on our own redeterminations. We give the
range of epochs used in Column 2 of Table \ref{table:SDSSHa};
we have, however, eliminated epoch MJD 56713 from all the light
curves as in many cases it was a clear outlier.
The time span used for each individual source was the
subset that gave the clearest results,
i.e., those with the smallest errors and/or the
least contamination by aliases.

\subsection{Fitting methodology}
In the remainder of this contribution,
we will examine the relationships among various
physical parameters via bivariate and multivariate
fits, first, to establish fundamental relationships
that will allow us to estimate central masses, and
second, to employ these relationships to develop 
predictive relationships to estimate the central masses.

We employ a fitting algorithm described by \citet{Cappellari13}
that combines the Least Trimmed Squares technique of
\citet{Rousseeuw06} and a least-squares fitting algorithm that
allows errors in all variables, as implemented in Paper I and
by \citet{DallaBonta18}. Most fits are bivariate fits of the
form
\begin{equation}
\label{eq:bivariate}
y = a + b(x-x_0), 
\end{equation}
where $x_0$ is the median value of the observable $x$. The fitting procedure minimizes the quantity
\begin{equation}
\label{eq:chi2line}
	\chi^2=\sum_{i=1}^N \frac{[a+b (x_i-x_0) - y_i]^2}
	{(b \Delta x_i)^2 + (\Delta y_i)^2 + \varepsilon_y^2},
\end{equation}
where $\Delta x_i$ and $\Delta y_i$ are the errors on the variables
$x_i$ and $y_i$, and $\varepsilon_y$ is the standard deviation
of the Gaussian
describing the distribution of intrinsic scatter in the $y$
parameter. The value of
$\varepsilon_y$ is adjusted iteratively so that the
$\chi^2$ per degree of freedom $\nu=N-2$ has the value of unity
expected for a good fit. The observed scatter is
\begin{equation}
\label{eq:Deltadef}
\Delta = \left\{ \frac{1}{N-2} \sum_{i=1}^N
\left[y_i - a - b\left(x_i - x_0\right) \right]^2 \right\}^{1/2}.
\end{equation}
The value of $\varepsilon_y$ is added in quadrature to the formal error when $y$ 
is used as a proxy for $x$.

As in Paper I, bivariate fits are intended to establish the physical relationships
among the various parameters and to fit residuals, as described below.
The initial mass estimation equations produced here
are based on multivariate fits of the general form
\begin{equation}
\label{eq:twoparameters}
z = a + b\left( x - x_0 \right) +c\left(y - y_0\right),
\end{equation}
where the parameters are as described above, plus an additional
observed parameter $y$ that has median value $y_0$.
Similarly to linear fits, the plane fitting minimizes the quantity
\begin{equation}
\label{eq:chi2plane}
	\chi^2=\sum_{i=1}^N \frac{[a + b (x_i-x_0) + c (y_i-y_0) - z_i]^2}
	{(b \Delta x_i)^2 + (c \Delta y_i)^2 + (\Delta z_i)^2 + \varepsilon_z^2},
\end{equation}
with $\Delta x_i$, $\Delta y_i$ and $\Delta z_i$ as the errors on the
variables $x_i,y_i,z_i$, and $\varepsilon_z$ as the sigma of the
Gaussian describing the distribution of intrinsic scatter in the $z$
coordinate; $\varepsilon_z$ is iteratively adjusted so that the $\chi^2$ per degrees of freedom $\nu=N-3$ has the value of unity expected for a good fit.
The observed scatter is
\begin{equation}
\label{eq:twoparmDeltadef}
\Delta = \left\{ \frac{1}{N-3} \sum_{i=1}^N
\left[{z_i} - a - b\left(x_i - x_0\right)
-c\left(y_i - y_0\right)\right]^2 \right\}^{1/2}.
\end{equation}

\section{Fits to the data}
\subsection{Fundamental relationships}
\begin{figure}
   \centering
   \resizebox{\hsize}{!}{\includegraphics{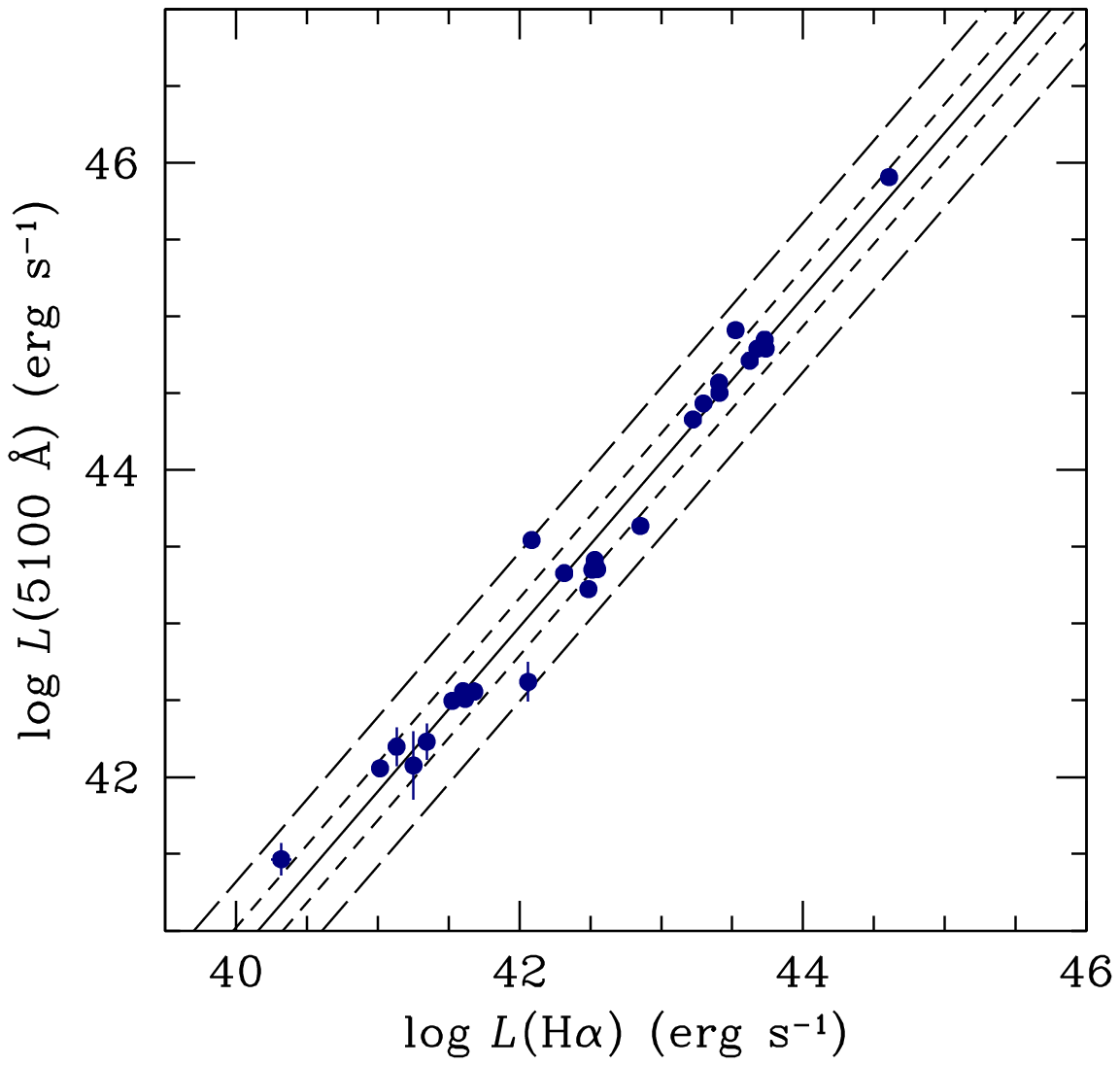}}
   \caption{Correlation between the luminosity of the
   broad H$\alpha$ emission line and the 
   starlight-corrected AGN continuum luminosity at 5100\,\AA.
   This figure includes only those AGNs with 
   host-galaxy starlight removal from the measured
   continuum based on {\em HST} high-resolution
   imaging \citep{Bentz13}, i.e., the AGNs
   listed in Table 1. The solid line represents
   the best fit to equation (\ref{eq:bivariate}) with
   parameters given in line 1 of Table \ref{table:fits1}.
   The short-dash lines show the $\pm 1 \sigma$ envelope
   and the long-dash lines show the $\pm 2.6\sigma$ 
   (99\% confidence level) envelope.}
         \label{figure:LHaL5100}%
    \end{figure}

One of the important results of Paper I is confirmation
of the tight relationship between the luminosity of the
broad H$\beta$ emission line with that of the AGN 
continuum. This is important as it eliminates the
necessity of quantifying the contribution of
contaminating starlight to the observed continuum flux and also avoids
possible complications from a contribution to the
continuum from a jet\footnote{We note, however, that 
only 3C\,273 = PG\,1226+023 and
RMID\,017 = SBS\,1411+533 are flat-spectrum radio quasars; 3C\,390.3
is also a radio source, but the jet is inclined to our line of sight.}.
This is even more critical in the H$\alpha$ region of
the spectrum where the starlight contamination is greater.
Figure \ref{figure:LHaL5100} shows the relationship
between the H$\alpha$ luminosity and the AGN
luminosity at 5100\,\AA\ (taken from Paper I).
The best fit parameters for this relationship are
given in line 1 of Table \ref{table:fits1}. The fit to this 
relationship shows that the luminosity of H$\alpha$ can
be used as a proxy for the AGN continuum at 5100\,\AA,
which itself is tacitly used as a proxy for the 
AGN ionizing continuum, as is the case with H$\beta$.

\begin{figure}
   \centering
   \resizebox{\hsize}{!}{\includegraphics{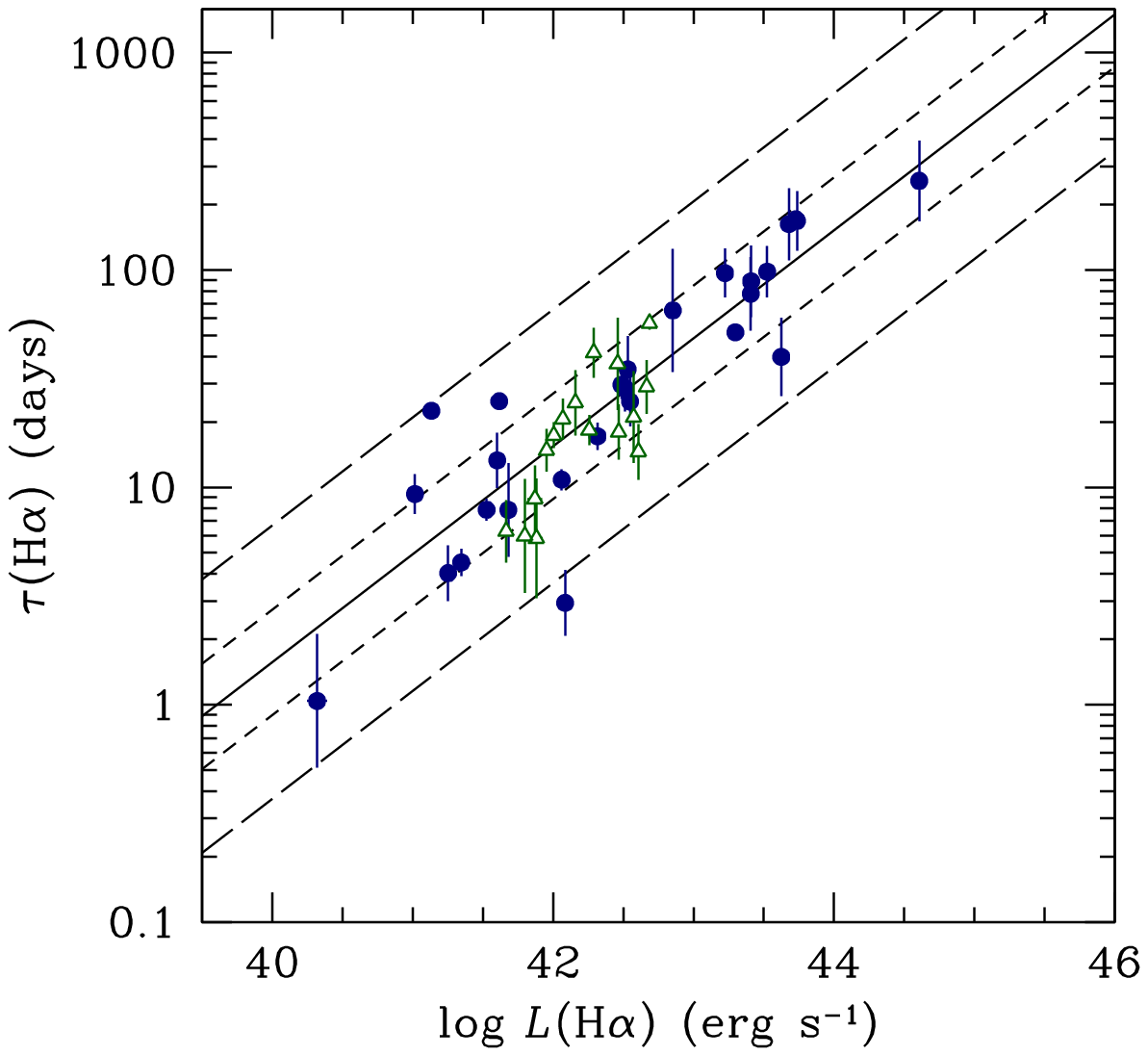}}
   \caption{Time-delayed response of the broad H$\alpha$
   line as as a function of    H$\alpha$ luminosity. 
   Since the response time is directly related to the
   broad-line region radius by $R=c\tau$, this is known
   as the $R$--$L$ relationship.
   Blue circles are from the reverberation-mapping
   database (Table \ref{table:RMDBHa}) and green triangles are from the SDSS project (Table \ref{table:SDSSHa}).
   The solid line shows the best fit to
   equation (\ref{eq:bivariate}) with parameters given
   in the second line of Table \ref{table:fits1}.
   The short- and long-dashed line show the
   $\pm1\sigma$ and $\pm2.6\sigma$ envelopes, respectively.
   }
   \label{figure:LHaLagHa}%
    \end{figure}

Reverberation-based black-hole masses (equation \ref{eq:virial})
are based on the measured lag $\tau$ of the emission-line flux variations 
relative to those of the continuum. Estimates of black-hole
masses based on individual spectra (``single epoch'' or SE masses)
are enabled by the well-known correlation between BLR radius $R = c\tau$ and AGN luminosity
\citep[][and additional historical references in Paper I]{Kaspi00, Kaspi05, Bentz06, Bentz09a, Bentz13}.
Figure \ref{figure:LHaLagHa} shows the relationship between the H$\alpha$ lag and luminosity
based on the data presented in Tables \ref{table:RMDBHa} and \ref{table:SDSSHa}. The best
fit parameters to these data are given in line 2 of Table \ref{table:fits1}; the slope of the relationship
is nearly exactly the canonical value $b = 0.5$.
This fundamentally establishes justification for seeking a single-epoch
predictor based on the H$\alpha$ line.

\begin{table*}
 \caption{Radius--Luminosity, Luminosity--Luminosity,
 and Line-Width Relations: $y = a + b(x-x_0)$}
\label{table:fits1}
 \centering
 \begin{tabular}{cllcccccc}
 \hline\hline
 Line & $x$ & $y$ & $a \pm \Delta a$ & $b \pm \Delta b$ &
 $x_0$ & $\varepsilon_y$ & $\Delta$ & Figures \\
 (1) & (2) & (3) & (4) & (5) & (6) & (7) & (8) & (9) \\
 \hline
1 & $\log L({\rm H}\alpha)$ &
$\log L_{\textrm{AGN}} (5100\,{\textrm \AA})$ &
$43.530 \pm 0.036$ &$1.072 \pm 0.036$ & 42.513 &
$0.174 \pm 0.036$ & 0.187 & \ref{figure:LHaL5100} \\
2 & $\log L({\mathrm H}\alpha)$ & $\log \tau({\mathrm H}\alpha$) &
$1.346 \pm 0.036$ & $ 0.497 \pm 0.016$ & 42.316 &
$0.206 \pm 0.036$ & 0.242 & \ref{figure:LHaLagHa} \\
3 & $\log \sigma_{\mathrm M}({\mathrm H}\alpha)$ &
 $\log \sigma_{\mathrm R}({\mathrm H}\alpha)$ & 
 $3.185 \pm 0.010$ & $1.074 \pm 0.053$ & 3.227 &
 $0.058 \pm 0.010 $ & 0.065 & \ref{figure:windowhawidths}(a)\\
4 & $\log {\rm FWHM}_{\rm M}({\rm H}\alpha)$ &
 $\log \sigma_{\mathrm R}({\mathrm H}\alpha)$ & 
 $3.205 \pm 0.025$ & $0.699 \pm 0.116$ & 3.511 &
 $0.140 \pm 0.023$ & 0.140 & \ref{figure:windowhawidths}(b)\\
 5 & $\log \mu_{\textrm{RM}}({\rm H}\alpha)$ &
$\log \mu_{\textrm{RM}}({\rm H}\beta)$ &
$7.049 \pm 0.049$ & $0.917 \pm 0.078$ & 6.956 &
$0.081 \pm 0.076$ & 0.289 & \ref{figure:VPRM} \\
  \hline
\end{tabular}
\end{table*}

The other parameter needed to compute a reverberation-based mass is the
emission-line width $\Delta V$ (equation \ref{eq:virial}). Broad emission lines
are typically comprised of multiple components, and the line-width measured used
in equation (\ref{eq:virial}) should be based only on the emission-line components
that are responding to the continuum variations. To isolate the variable part 
of the emission line, a root-mean-square residual spectrum (for brevity hereafter referred to as the ``RMS spectrum'')
is constructed. The mean spectrum is defined by
\begin{equation}
\label{eq:meandef}
    \overline{F}(\lambda) = \frac{1}{N} \sum^N_1 F_i(\lambda)
\end{equation}
where $F_i(\lambda)$ is flux of the $ith$ spectrum of the time series at wavelength $\lambda$
and $N$ is the total number of spectra. The RMS spectrum is then defined by
\begin{equation}
    \label{eq:RMSdef}
    \sigma_{\rm rms}(\lambda) = \left\{ \frac{1}{N-1} \sum^N_{1} \left[ F_i(\lambda) - 
    \overline{F}(\lambda) \right]^2 \right\}^{1/2}.
\end{equation}

\begin{figure*}
   \centering
   \resizebox{\hsize}{!}{\includegraphics{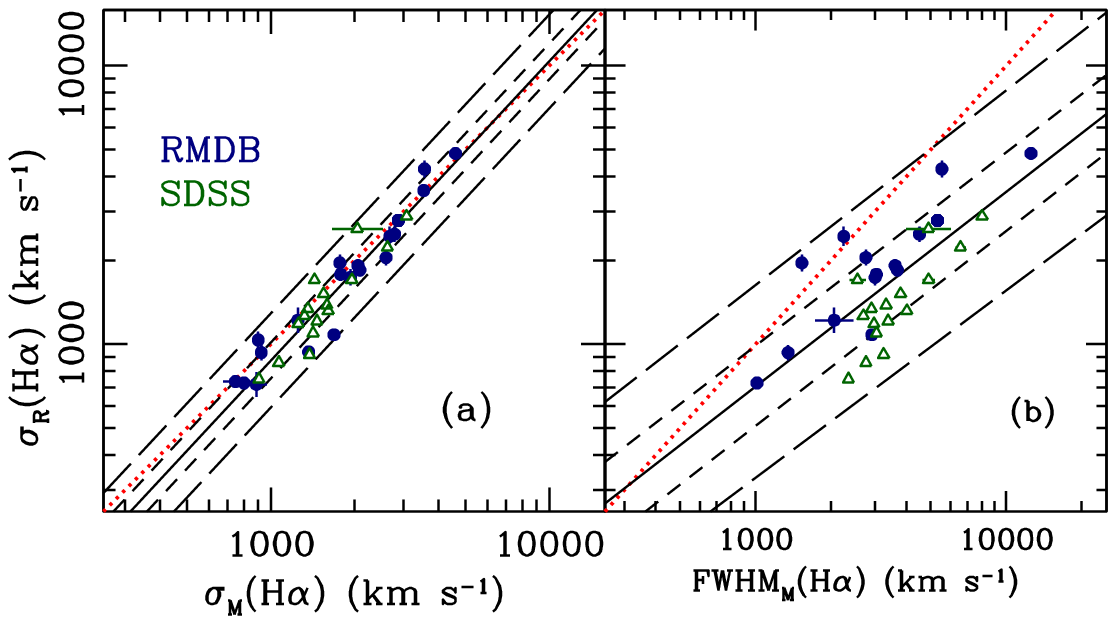}}
   \caption{(a) Relationship between H$\alpha$ line dispersion in the
   RMS spectrum $\sigma_{\mathrm R}({\rm H}\alpha)$
   and the mean spectrum $\sigma_{\mathrm M}({\rm H}\alpha)$.
   (b) Relationship between H$\alpha$ line dispersion in the
   rms spectrum $\sigma_{\mathrm R}({\rm H}\alpha)$ and FWHM
   in the mean spectrum ${\rm FWHM}_{\mathrm M}({\rm H}\alpha)$.
   Blue points are from Table 1 (RMDB sample), green are from Table 2 (SDSS sample).The solid lines are the best fit to equation (\ref{eq:bivariate})
   with coefficients from Table \ref{table:fits1}. 
   The short- and long-dashed lines
   indicate the $\pm 1 \sigma$ and $\pm 2.6 \sigma$ envelopes. The red dotted lines indicate where the two measures are equal.}
      \label{figure:windowhawidths}%
    \end{figure*}

There are multiple parameters that might be used to characterize the emission-line width.
Most commonly used are the full-width at half maximum (FWHM) and the line
dispersion, defined by
\begin{equation}
    \label{eq:sigmadef}
    \sigma_{\rm line} =\left[ \frac{\int (\lambda - \lambda_0)^2 P(\lambda)\,d\lambda}
    {\int P(\lambda)\,d\lambda} \right]^{1/2},
    \end{equation}
where $P(\lambda)$ is the  line profile and $\lambda_0$ is the line centroid
\begin{equation}
\label{eq:centroid}
\lambda_0 = \frac{\int \lambda P(\lambda)\,d\lambda}{\int P(\lambda)\,d\lambda}.
\end{equation}
Paper I presents detailed arguments that the line-dispersion in the RMS spectrum
$\sigma_{\rm R}$ is better than
FWHM in the RMS spectrum, FWHM$_{\rm R}$  for computing reverberation masses. We have also carried out a preliminary
investigation of other line-width measures and find that there
are other good proxies for $\sigma_{\rm R}$
\citep{EDB22}, but this discussion is beyond the scope
of the current work and will be pursued elsewhere.
The aim here is then to determine, given a single spectrum, what line-width measure in the {\it mean} or a {\it single}
spectrum (since the mean spectrum
is a reasonable representation of a single-epoch spectrum in the time series) is the better proxy for $\sigma_{\rm R}$. Figure \ref{figure:windowhawidths}a
shows the relationship between $\sigma_{\rm R}$ and
the line-dispersion in the mean spectrum $\sigma_{\rm M}$.
Figure \ref{figure:windowhawidths}(b) shows the relationship
between $\sigma_{\rm R}$
and FWHM in the mean spectrum ${\rm FWHM}_{\rm M}$.
Best fit relationships between pairs of parameters are given in third and
fourth lines of Table \ref{table:fits1}.
As is the case with H$\beta$ as described in Paper I, $\sigma_{\rm M}$ is an
excellent proxy for $\sigma_{\rm R}$. On the other hand, 
${\rm FWHM}_{\rm M}$
can also be used as a proxy for $\sigma_{\rm R}$, but the relationship is
not close to linear and the additional uncertainty introduced ($\varepsilon_y$) is more than
twice as large as that introduced by $\sigma_{\rm M}$.

\begin{figure}
   \centering
   \resizebox{\hsize}{!}{\includegraphics{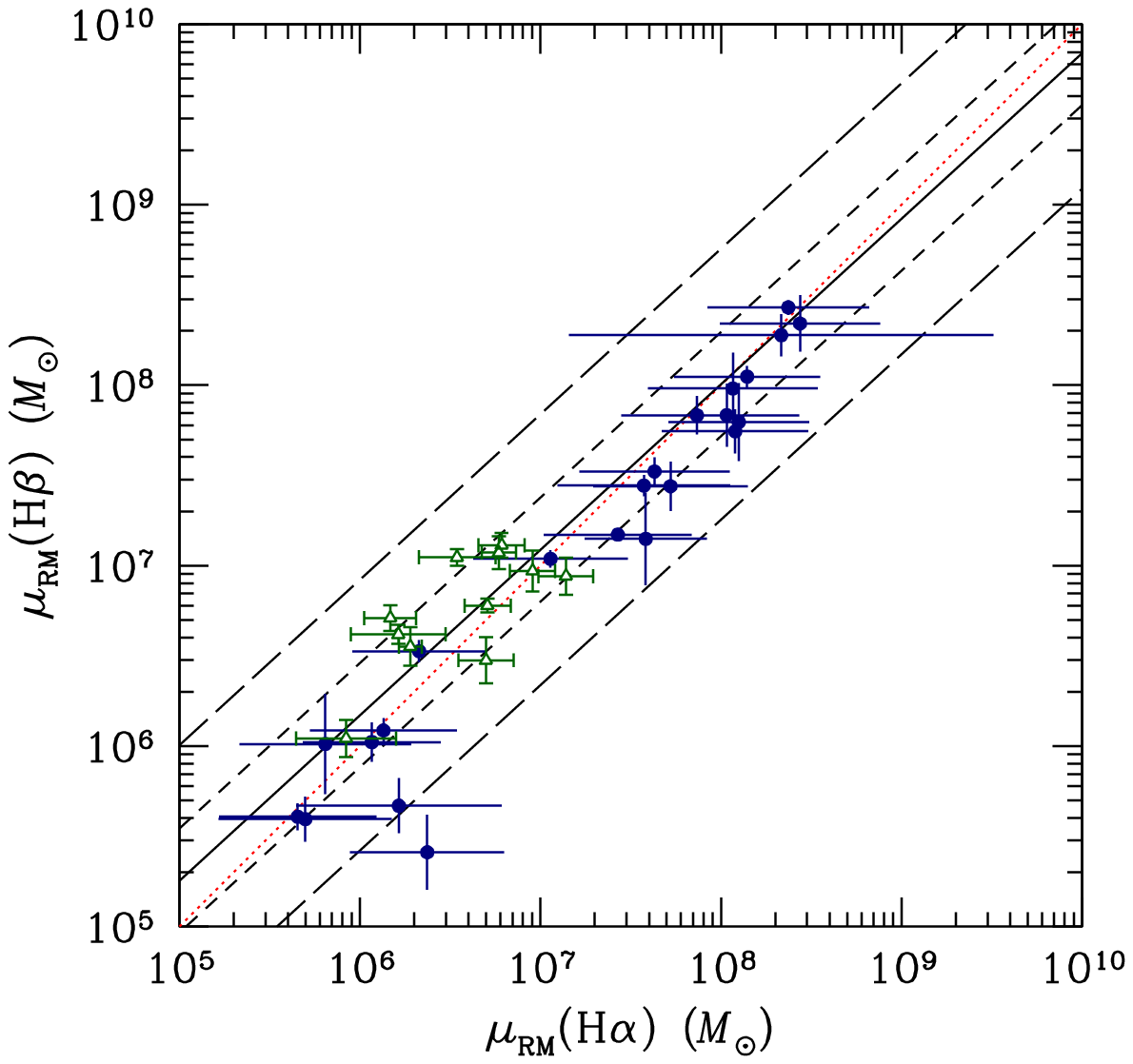}}
   \caption{Comparison between virial products
   based on the H$\alpha$ data presented in Tables
   \ref{table:RMDBHa} (blue circles) and 
   \ref{table:SDSSHa} (green triangles)
   and virial products based on H$\beta$ in Paper I.
   The dashed red line shows the locus where
   the two values are equal. The solid black line 
   shows the best fit to the data. The short- and long-dashed
   lines show the $\pm1 \sigma$ and $\pm2.6\sigma$ envelopes,
   respectively.
   The largest 
   outlier is Mrk\,202, which \cite{Bentz10} flag as having an
   especially dubious H$\alpha$ lag measurement.
   }
   \label{figure:VPRM}%
    \end{figure}

At this point, we compare 
in Figure \ref{figure:VPRM}
the virial products obtained
with the H$\alpha$ data in Tables \ref{table:RMDBHa} and \ref{table:SDSSHa}
with the H$\beta$-based virial products we obtained in Paper I.
For individual sources, in most cases the two virial products agree to
within the uncertainties.
A simple fit to this distribution, with
resulting coefficients shown in line 5 of
Table \ref{table:fits1}, confirms that the slope is
less than unity and that the H$\alpha$-based
virial product sightly exceeds the H$\beta$-based
values with increasing mass.
In the analysis that follows, we use the H$\beta$-based masses as our
reference because the typical uncertainties ($\sim 0.113$\,dex) are
considerably smaller than those associated with the H$\alpha$-based masses
($\sim0.358$\,dex). 

\subsection{Fits and corrections}
The correlations identified above justify a search for a single-epoch formula
to estimate black hole masses from H$\alpha$. As a first approximation,
we begin by trying to reproduce the H$\beta$ RM virial product
with the expectation that the BLR radius can be determined from
the luminosity and that the line width in the mean spectrum
can be used as a proxy for $\sigma_{\rm R}$.
The following equations are used:
\begin{eqnarray}
\label{eq:sigmainitial}
\log \mu_{\rm RM}({\rm H}\beta) & = &a + b\left[ \log L({\rm H}\alpha) - x_0\right] \nonumber \\
& &+ c\left[\log \sigma_{\rm M}({\rm H}\alpha) - y_0\right]
\end{eqnarray}
and 
\begin{eqnarray}
\label{eq:FWHMinitial}
\log \mu_{\rm RM}({\rm H}\beta) & = & a + b\left[ \log L({\rm H}\alpha) - x_0\right] \nonumber \\
& &+ c\left[\log {\rm FWHM}_{\rm M}({\rm H}\alpha) - y_0\right].
\end{eqnarray}
The best fits to these equations are given in Table \ref{table:multivariate}.
These can be used to produce initial single-epoch (SE) predictors
\begin{eqnarray}
\label{eq:SEsigm}
\log \mu_{\rm SE}({\rm H}\alpha) & = & 6.996 + 0.501\left[\log L({\rm H}\alpha) -42.267\right] 
\nonumber \\
& & + 2.397\left[\log \sigma_{\rm M}({\rm H}\alpha) - 3.227\right]
\end{eqnarray}
and 
\begin{eqnarray}
\label{eq:SEfwm}
\log \mu_{\rm SE}({\rm H}\alpha) & = & 7.082 + 0.583\left[\log L({\rm H}\alpha) -42.531\right] 
\nonumber \\
& & + 1.173\left[\log {\rm FWHM}_{\rm M}({\rm H}\alpha) - 3.314\right].
\end{eqnarray}

\begin{figure*}
   \centering
   \resizebox{\hsize}{!}{\includegraphics{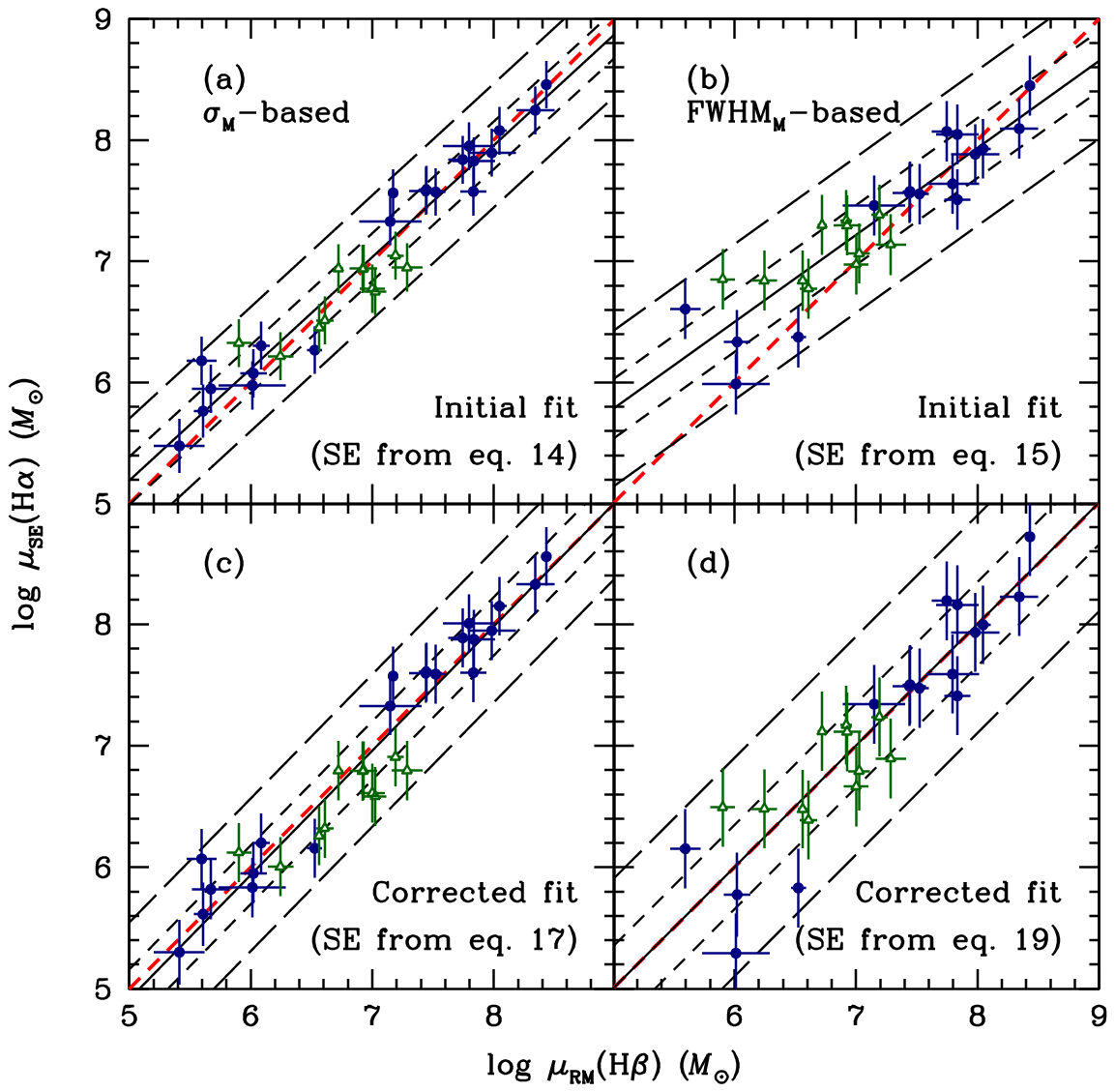}}
   \caption{Upper: Single-epoch H$\alpha$-based virial product predictions
   using equations (\ref{eq:SEsigm}) and (\ref{eq:SEfwm})
   in panels (a) and (b) respectively. The coefficients for the best
   fit appear in the first two lines of Table \ref{table:fits1}.
   Blue circles are data from Table \ref{table:RMDBHa} and
   green triangles are from Table \ref{table:SDSSHa}.
   The solid line is the best fit to the data and the red dotted
   line shows where the measures are equal. The short- and long-dashed
   lines show the $\pm1\sigma$ and $\pm 2.6\sigma$ envelopes, respectively.
   Lower: Corrected single-epoch masses from H$\alpha$-based
   virial product predictions using equation
   (\ref{eq:Hamass_sigma}) (panel c) and equation
   (\ref{eq:Hamass_FWHM}) (panel d) in both cases with 
   $\log f = 0$ arbitrarily.   
   }
    \label{figure:windowvp}%
    \end{figure*}

\begin{figure*}
   \centering
   \resizebox{\hsize}{!}{\includegraphics{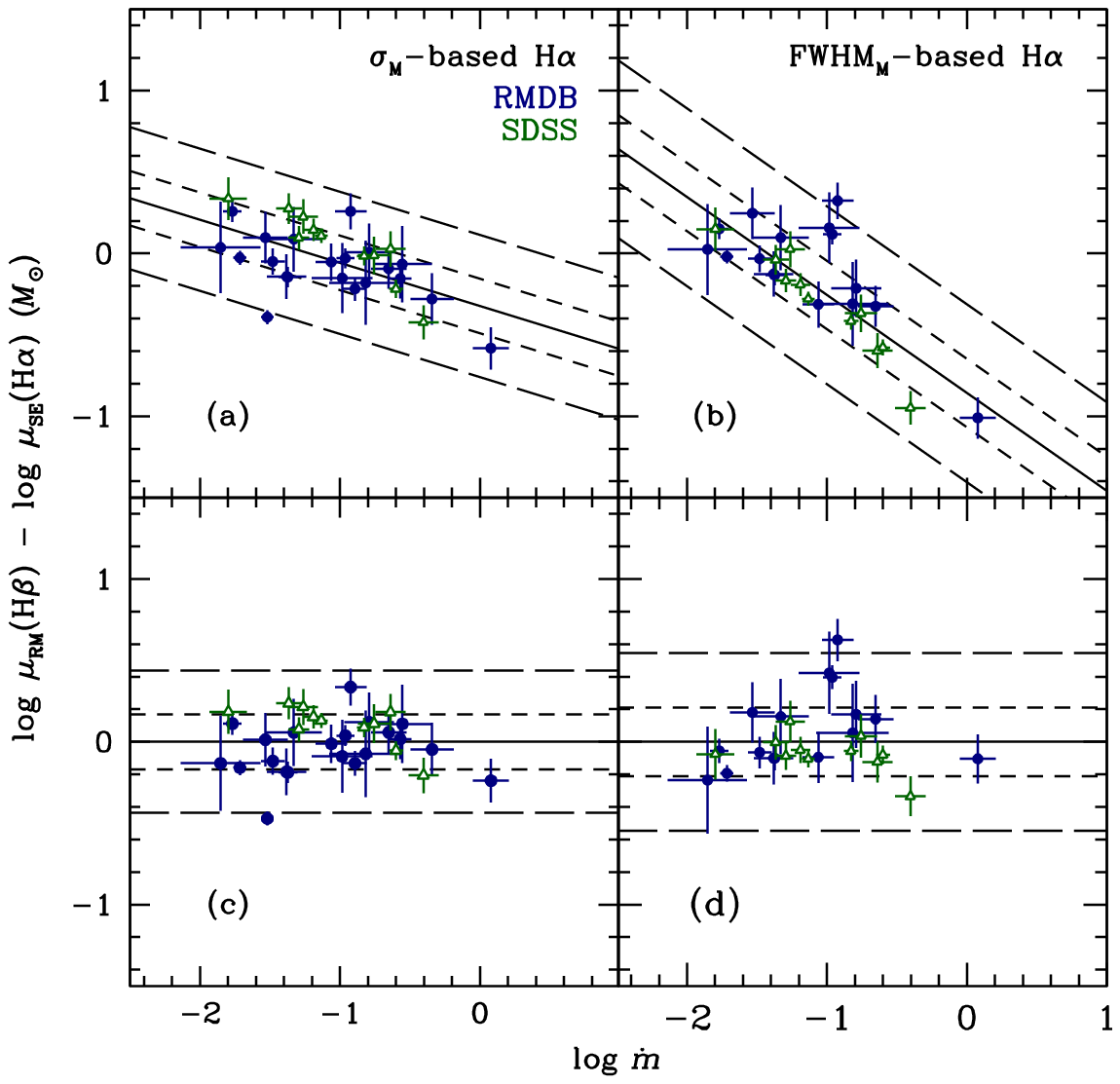}}
   \caption{(a) Mass residuals (equation \ref{eq:Eddcor}) are the difference between the
   measured reverberation virial products and those 
   predicted by equation (\ref{eq:sigmainitial}). The residuals are plotted
   vs.\ Eddington ratio. The solid line represents the 
   best fit, the short-dashed lines represent the
   $\pm 1 \sigma$ envelope and the long-dashed line
   represent the $\pm 2.6 \sigma$ envelope. Blue circles
   are from Table \ref{table:RMDBHa} and green triangles are
   from Table \ref{table:SDSSHa}.
   (b) Mass residuals are the difference between the 
   measured reverberation virial products and those predicted by
   equation (\ref{eq:SEfwm}). 
   Panels (c) and (d) show residuals
   after subtraction of the best fit relations shown in
   panels (a) and (b).}
 \label{figure:windowharesiduals}%
    \end{figure*}

These data and best fits to them are shown in Figure \ref{figure:windowvp},
equation (\ref{eq:SEsigm}) in panel (a) and equation (\ref{eq:SEfwm}) in panel (b).
In both cases, the slope $b$ is shallower than unity, indicating that 
the line luminosity and line width are, by themselves, unable to accurately
predict the reverberation measurement $\mu_{\rm RM}$. 
As noted above, in Paper I, we found that
the residuals in this relationship were well-correlated with
Eddington ratio, which is the ratio of actual mass-accretion
rate relative to the maximum or Eddington rate. This finding
is in agreement with the conclusions of others who have
investigated the $R$--$L$ relationship
\citep[e.g.,][]{Du16, Du18, Grier17b, Du19, Martinez19, Alvarez20}.
In the upper panels of Figure \ref{figure:windowharesiduals}, we show the residuals in
the $\mu_{\rm RM}$--$\mu_{\rm SE}$ relationship for
equations (\ref{eq:SEsigm}) (panel a)  and (\ref{eq:SEfwm}) (panel b).
As in Paper I, we compute a correction to the SE mass by fitting the
relationship
\begin{equation}
\label{eq:Eddcor}
\Delta \log \mu = \log \mu_{\rm RM} - \log \mu_{\rm SE} =
a + b (\log \dot{m} - x_0).
\end{equation}
Our assumptions and calculations of the Eddington ratio,
the most important assumption being our use of the
bolometric correction from \cite{Netzer19},
are given in Paper I with the single modification that
we use equation (\ref{eq:bivariate}) with the relationship
shown in Figure \ref{figure:LHaL5100} to
substitute $L({\rm H}\alpha$) for $L_{\textrm{AGN}}(5100\,{\textrm \AA})$.
The reason our correction works is because the simple
assumptions we make to compute the Eddington ratio
depend only on $L({\rm H}\alpha)$ 
(or equivalently, $L_{\textrm{AGN}} (5100\,{\textrm \AA})$ or $L({\textrm H}\beta)$) and
$\mu_{\rm RM}$, which are known for this sample.
The best-fit parameters for equation (\ref{eq:Eddcor}) are
given in lines 3 and 4 of Table
\ref{table:fits2}. The bottom panels of Figure \ref{figure:windowharesiduals} show the effect of this correction
on the residuals.

Applying the correction of equation (\ref{eq:Eddcor}) to the
single-epoch masses in the top panels of Figure \ref{figure:windowvp}
yields the corrected single-epoch masses shown in the bottom
panels of the same figure. The best-fit parameters for the revised relationship
are given in lines 5 and 6 of Table \ref{table:fits2} for the case of
$\sigma_{\rm M}$ and ${\rm FWHM}_{\rm M}$-based masses, respectively.
Note in particular that the slopes of these relationships are
very close to the expected value of unity, indicating that the 
three variables identified --- line luminosity, line width,
and Eddington ratio --- are sufficient to estimate the black hole mass to fairly high accuracy.


 \begin{table*}
\caption{Initial, Residual, and Final Fits: $y = a +b(x-x_0)$}
\label{table:fits2}
\centering
\begin{tabular}{lcllcccccc}
Line & Basis & $x$ & $y$ & $a + \Delta a$ & $b + \Delta b$ &
$x_0$ & $\varepsilon_y$ & $\Delta$ & Figures\\
(1) & (2) & (3) & (4) & (5) & (6)& (7) & (8) & (9) & (10) \\
\hline\hline
Initial: \\
1 & $\sigma_{\mathrm M}$ & $\log \mu_{\rm RM}$ & 
$\log \mu_{\rm SE} $  & $7.058\pm 0.036$ & $0.922 \pm 0.045$ &
7.026 & $0.174 \pm 0.035$ & 0.196 & \ref{figure:windowvp}(a)\\
2 & ${\rm FWHM_{\rm M}}$ & $\log \mu_{\rm RM}$ &
$\log \mu_{\rm SE}$ & 
$7.323 \pm 0.048$ & $0.718 \pm 0.064$ & 7.149 &$0.231 \pm 0.048$ 
& 0.247 & \ref{figure:windowvp}(b)\\
Residual: \\
3 & $\sigma_{\mathrm M}$ & $\log \dot{m}$ & $\Delta \log \mu$ &
$-0.062 \pm 0.033$ & $-0.265 \pm 0.072$ & $-0.985$ &
$0.143 \pm 0.030$ & 0.168 & \ref{figure:windowharesiduals}(a)\\
4 & ${\rm FWHM_{\mathrm M}} $ & $\log \dot{m}$ & $\Delta \log \mu$ &
$-0.175 \pm 0.040$ & $-0.602 \pm 0.090$ & $-1.134$ & 
$0.162 \pm 0.045$ & 0.210 & \ref{figure:windowharesiduals}(b) \\
Final: \\
5 & $\sigma_{\mathrm M}$ & $\log \mu_{\rm RM}$ & 
$\log \mu_{\rm SE} $ & $6.988 \pm 0.044$ & $1.022 \pm 0.055$ &
7.026 & $0.219 \pm 0.042$ & 0.241 & \ref{figure:windowvp}(c)\\
6 & ${\rm FWHM_{\rm M}}$ & $\log \mu_{\rm RM}$ &
$\log \mu_{\rm SE}$ & $7.151 \pm 0.067$ & $1.000 \pm 0.089$ & 7.149 &
$0.322 \pm 0.067$ & 0.345 & \ref{figure:windowvp}(d)\\
\hline
\end{tabular}
 \end{table*}


\section{Formulas for mass estimation}

Our initial estimates (equations \ref{eq:SEsigm} and \ref{eq:SEfwm}) can be combined
with the Eddington rate correction (equation \ref{eq:Eddcor}) which is inverted to solve for an
estimate of $\mu_{\rm RM}$ based solely on $L({\rm H}\alpha)$
and $\sigma_{\rm M}({\rm H}\alpha)$. This yields our equations for 
the corrected SE virial product
estimator, with zero-points
adjusted for convenience and 
for consistency with Paper I,

\begin{eqnarray}
\label{eq:Hamass_sigma}
\nonumber
    \log M_{\rm SE} &  =  & \log f + 7.413 + 
    0.554\left[\log L({\rm H}\alpha) - 42 \right] \\
    & & + 2.61\left[\log \sigma_{\rm M}({\rm H}\alpha) - 3.5 \right],
\end{eqnarray}
which has an associated uncertainty
\begin{eqnarray}
\label{eq:Hamass_sigma_error}
\nonumber
\Delta \log M_{\rm SE} &  =  & \left\{ \left( \Delta \log f \right)^2 +
\left[ 0.554\, \Delta \log L({\rm H}\alpha) \right]^2 \right. \\
& & \left.  + \left[ 2.61\,\Delta \log \sigma_{\rm M}({\rm H}\alpha) \right]^2 
\right\}^{1/2}.
\end{eqnarray}
Note that the intrinsic scatter, $\varepsilon_y = 0.219$, needs to be added in quadrature to the formal uncertainty.

Similarly, in the case where FWHM is used as the line-width measure, 
\begin{eqnarray}
\label{eq:Hamass_FWHM}
\nonumber
    \log M_{\rm SE} &  =  & \log f + 6.688 + 
    0.812\left[\log L({\rm H}\alpha) - 42 \right] \\
    & & + 1.634\left[\log {\rm FWHM}_{\rm M}({\rm H}\alpha) - 3.5 \right],
\end{eqnarray}
which has an associated uncertainty
\begin{eqnarray}
\label{eq:Hamass_FWHM_error}
\nonumber
\Delta \log M_{\rm SE} &  =  & \left\{ \left( \Delta \log f \right)^2 +
\left[ 0.812\, \Delta \log L({\rm H}\alpha) \right]^2 \right. \\
& & \left.  + \left[ 1.634\,\Delta \log {\rm FWHM}_{\rm M}({\rm H}\alpha) \right]^2 
\right\}^{1/2}.
\end{eqnarray}
Again, the intrinsic scatter, $\varepsilon_y =0.332$,
needs to be combined in quadrature with the
formal uncertainty.

The mean scale factor is determined by
calibrating the virial products $\mu_{\rm RM}$ to
the $M_{\rm BH}$--$\sigma_*$ relations.
Our adopted value, based on the most recent analysis of the
largest database,
is $\langle \log f \rangle = 0.683 \pm  0.150$ \citep{Batiste17}.
The error on the mean is $\Delta \log f = 0.030$\,dex and
this should also be folded into the mass 
estimate uncertainty.

\section{Discussion}

\subsection{Limitations}
The SE mass predictors developed here and in Paper I
have included sources in the luminosity range
\begin{equation}
\nonumber
41 \la \log L_{\rm AGN}(5100\,{\rm \AA})\ ({\rm erg\,s}^{-1}) \la 46
\end{equation}
for the Balmer lines and
\begin{equation}
\nonumber
39.5 \la L(1350\,{\AA})\ ({\rm erg\,s}^{-1}) \la 47
\end{equation}
for \ion{C}{IV}, though the sample size for the latter is
very poor below $\log L(1350\,{\rm \AA}) \approx 42$
(e.g., Figure 9 of Paper I). This does, however, cover
most of the known range of AGN activity.
The range of Eddington ratio covered by these estimators
is
\begin{equation}
\nonumber
    -2 \la \log \dot{m} \la 0
\end{equation}
for the Balmer lines with extension to lower rates
in \ion{C}{IV}, as low as $\log \dot{m} \approx -3$,
but with poor sampling.
This reaches close to the lowest Eddington ratios
expected for broad-line AGNs. Extension to super-Eddington
rates (i.e., $\dot{m} > 1$) remains to be explored.

\subsection{On the importance of line-width measures}

Much of this work has been focused on how the
line-width measures are used. In particular, we have
argued that if FWHM is used as $\Delta V$ in
equation (\ref{eq:virial}), the mass scale will be erroneously
stretched. 
At larger line widths (higher mass at fixed
luminosity), the ratio ${\rm FWHM}/\sigma$ is high so that
the higher masses are overestimated by using FWHM. Similarly,
for narrower lines (lower masses at fixed luminosity)
${\rm FWHM}/\sigma$ is low and thus lower masses
are consequently underestimated by using FWHM.
This point was
made very clear by \cite{Rafiee11} who demonstrated this
for fixed intervals of luminosity. The key point is that
the mass scale is stretched {\em at fixed luminosity.}

\begin{figure}
   \centering
   \resizebox{\hsize}{!}{\includegraphics{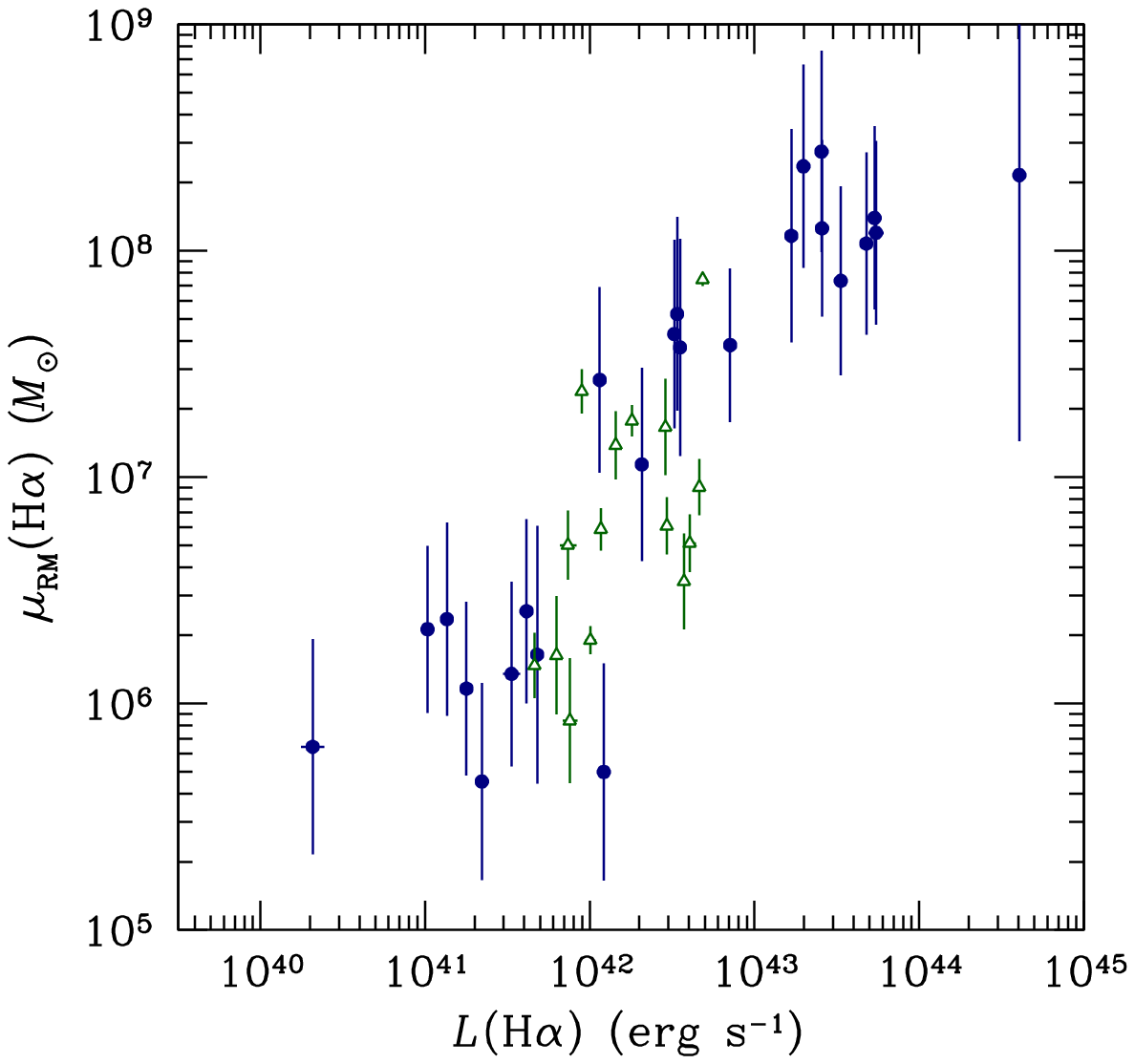}}
   \caption{Correlation between virial product
   and luminosity of the H$\alpha$ emission line.
   The apparent correlation between mass and luminosity is in fact due to selection effects. Sources in the lower right (high luminosity, low mass) are generally excluded by the Eddington limit. Sources in the upper left (low luminosity, high mass) are scarce (a) because high mass objects are rare, therefore mostly at large distances and therefore faint, and (b) because their accretion rates are so low that they do not manifest themselves as AGNs.
  }
   \label{figure:ML}%
    \end{figure}

This may obscure the fact that the single most important
parameter in AGN black-hole mass estimation is
luminosity. This is because the range of luminosity
(over four orders of magnitude in the sample 
discussed here) is much larger than the range of
line widths (spanning about a single order of magnitude
in this sample). Figure \ref{figure:ML} suggests that
there is in fact a correlation between
luminosity and mass and a crude mass estimate
can be made based on luminosity alone, which is tantamount
to assuming that the range of Eddington ratio $\dot{m}$ is
very narrow; indeed this
realization led to a suggestion that
the line width is superfluous and contains
little if any additional leverage in estimating 
AGN black hole masses \citep{Croom11}. This is true only if
the line-width measurements are very inaccurate
(as they sometimes are if they are measured from
survey-quality data) or if
one is willing to settle for a less than order-of-magnitude
mass estimate. More importantly, however, 
one must be cognizant of the 
fact that selection effects militate against identification
of sources in the upper left and lower right parts of this Figure.
Luminosity alone is simply not a very good predictor
of black hole mass.

\subsection{Comparison with GRAVITY results}
\begin{figure}
   \centering
   \resizebox{\hsize}{!}{\includegraphics{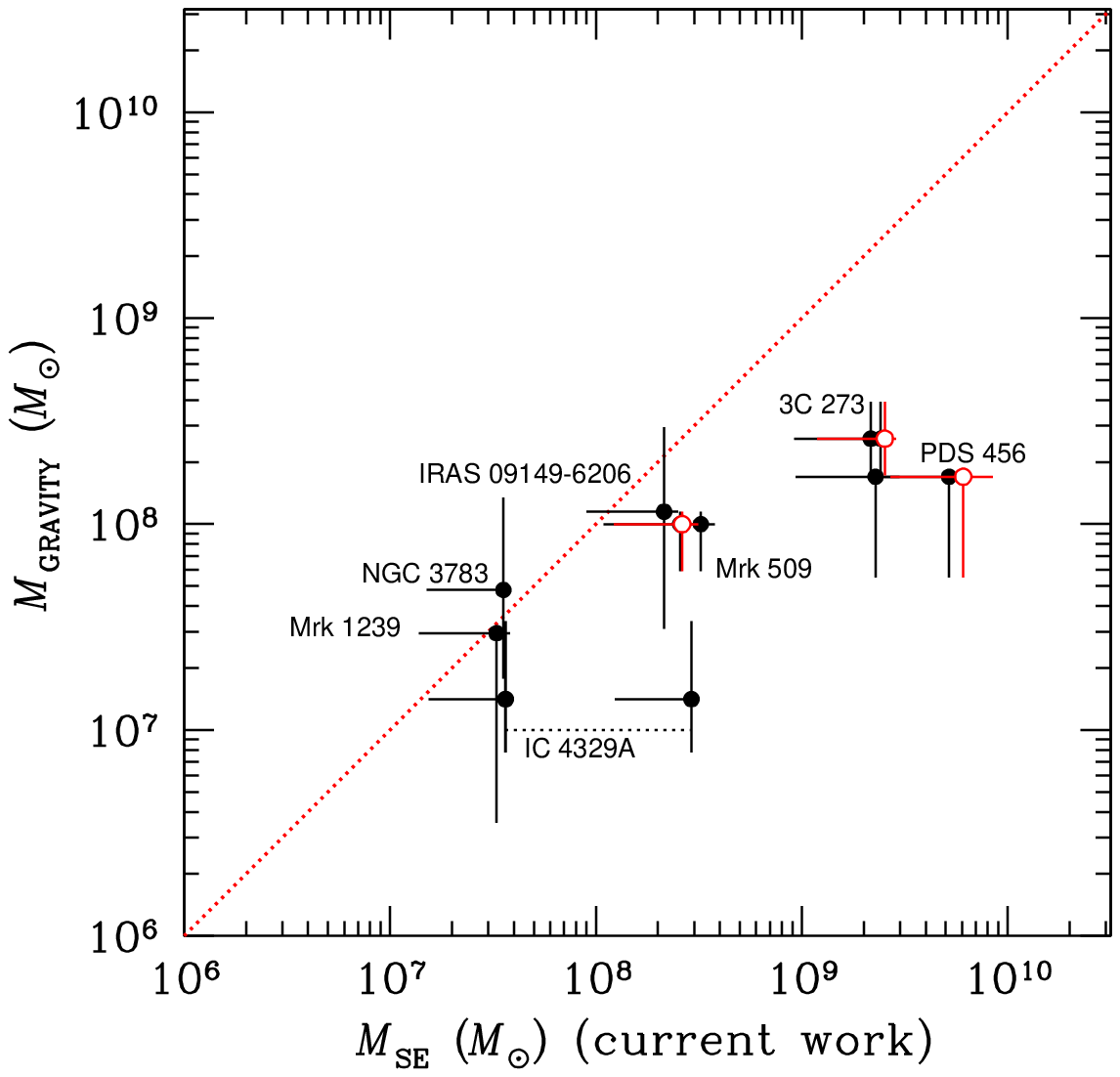}}
   \caption{Comparison of the mass predictions
   from GRAVITY and SE estimates from the current work.
   H$\beta$-based SE masses are in black, H$\alpha$-based
   are in red.
    {\it Note that this figure plots mass
   rather than virial product.}
   }
    \label{figure:gravitycomp3}
    \end{figure}

As noted earlier, there are a handful of sources that have
been spatially resolved with the GRAVITY interferometer and
have yielded mass measurements. Here we take luminosity
and line width measures from the published literature
and use equation (\ref{eq:Hamass_FWHM})
and equation (40) from Paper I to make comparisons between
the GRAVITY measurements and our SE predictors. The results
are summarized in Table \ref{table:GRAVITY}
and shown in Figure \ref{figure:gravitycomp3}. 
Note that we have assumed
$\log f = 0.683$. 

Figure \ref{figure:gravitycomp3} shows that the
GRAVITY and SE-based masses are generally consistent
at the low-mass end, but not at the high-mass end.
In the case of IC\,4329A, one SE-based prediction is
considerably larger than the other two; this is because
the estimate from \citet{Li24} assumes $\sim 2$\, mag
of internal extinction of the nucleus. The actual
reverberation measurement, using measurements from
\citet{Li24}, equation (A1) from Paper I,
and $\log f =0.683)$ is $\sim 7.75$ in log solar units,
closer to the other SE measurements than the
estimate based on an internal extinction correction,
which suggests that this correction is too large.
For the two highest mass sources, 3C\,273 and
PDS\,456, the GRAVITY and SE 
masses are in poorer agreement, and we note that in
both cases, na\"{\i}ve application of the Eddington
limits suggests both masses should exceed
$\sim10^9\,M_\odot$
\citep[e.g.,][]{Nardini15}.
In general, the SE estimates are in better agreement
with the RM measurements than the GRAVITY measurements, 
when they are available.

\begin{table*}
\caption{Comparison of Single Epoch (SE) Estimates
with GRAVITY Meaurements}
\label{table:GRAVITY}
\centering
\begin{tabular}{lccccc}
 & $\log (M/M_\odot) $ & & 
$\log (M/M_\odot) $   &  \\
Source & (GRAVITY) & Ref & (SE) & Line & Ref \\
(1) & (2) & (3) & (4) & (5) & (6)\\
\hline\hline
IRAS\,09149-6206 & $8.06^{+0.41}_{-0.57}$ & 1 &
$8.331 \pm 0.377$ & H$\beta$ & 5 \\
Mrk\,1239 & $7.47 \pm 0.92$ & 2 &
$7.518 \pm 0.377$ & H$\beta$ & 6 \\
NGC\,3783 & $7.68^{+0.45}_{-0.43}$ & 3 &
$7.551 \pm 0.372$ & H$\beta$ & 7 \\
3C\,273 & $8.41 \pm 0.18$ & 4 & 
$9.403 \pm 0.328$ & H$\alpha$ & 8 \\
& & & $9.383 \pm 0.372$ & H$\beta$ & 8 \\
& & & $9.335 \pm 0.372$ & H$\beta$ & 9 \\
IC\,4329A & $7.15^{+0.38}_{-0.26}$ & 2 &
$7.563 \pm 0.373$ & H$\beta$ & 10 \\
& & & $7.560 \pm 0.372$ & H$\beta$ & 11,12 \\
& & & $8.484\pm0.372$ & H$\beta$ & 6 \\
PDS\,456 & $8.23^{+0.01}_{-0.49}$ & 2 &
$9.784 \pm 0.354$ & H$\alpha$ & 13\\
& & & $9.358 \pm 0.389$ & H$\beta$ &13,14\\
& & & $9.715\pm0.373$ & H$\beta$ & 6 \\
Mrk\,509 & $8.00^{+0.06}_{-0.23}$ & 2 &
$8.419 \pm 0.333$ & H$\alpha$ & 15 \\
& & & $8.510 \pm 0.377$ & H$\beta$ & 15 \\
& & & $8.409 \pm 0.373$ & H$\beta$ & 6\\
\hline
\end{tabular}
\tablebib{
       (1)~\citet{GRAVITY20}; (2)~\citet{GRAVITY24}; (3)~\citet{GRAVITY21a}; (4)~\citet{GRAVITY18};
       (5)~\citet{Perez89}; (6)~\citet{Li24};
       (7)~\citet{Bentz21}; (8)~\citet{Kaspi00};
       (9)~\citet{Zhang19}; (10)~\citet{Bentz23};
       (11)~\citet{Winge96}; (12)~\citet{Collin06};
       (13)~\citet{Simpson99}; (14)~\citet{Torres97};
       (15)~\citet{Osterbrock77}.
        }
 \end{table*}

\subsection{Comparison with other single-epoch estimates}
\begin{figure}
   \centering
   \resizebox{\hsize}{!}{\includegraphics{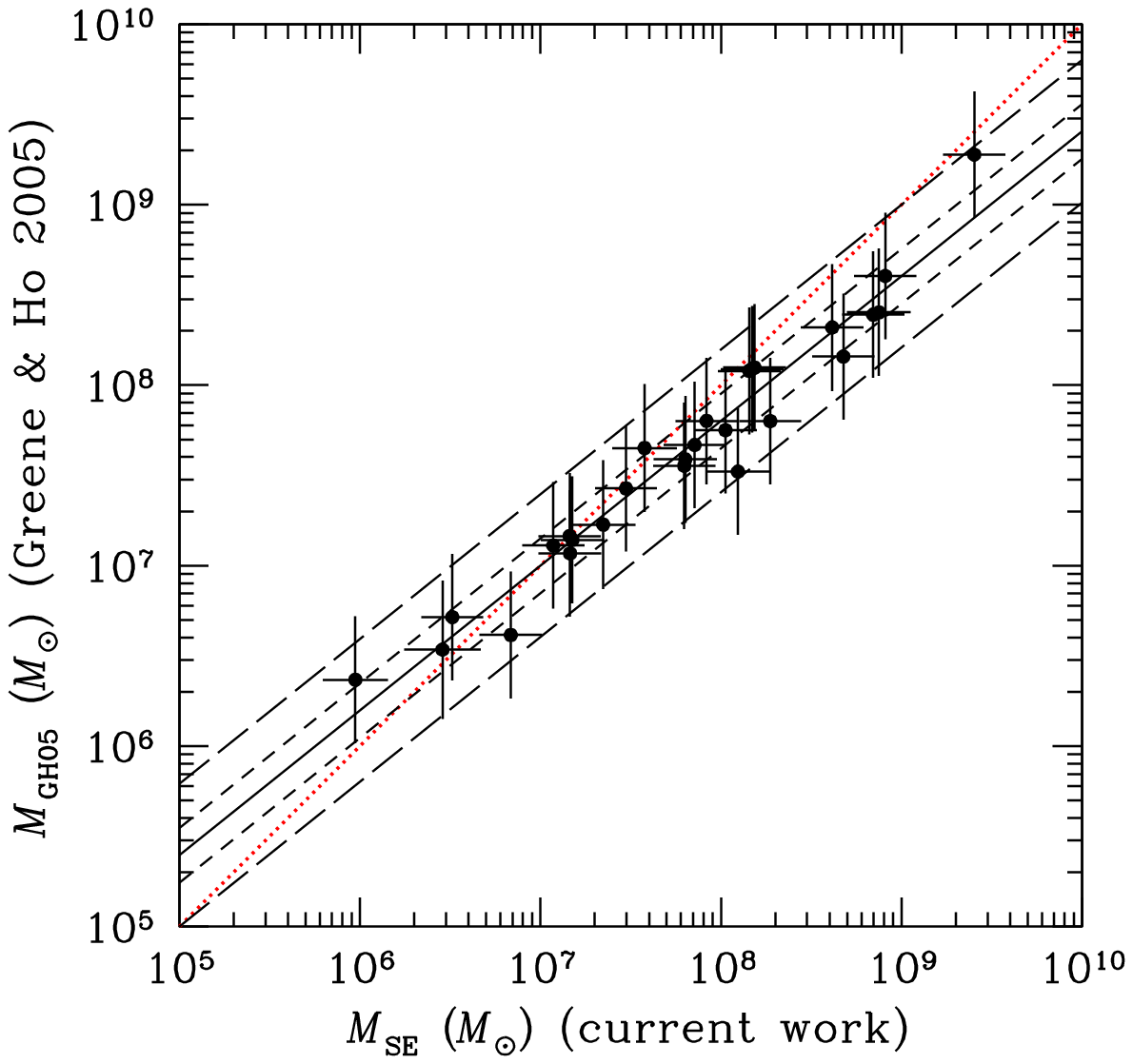}}
   \caption{Direct comparison of the mass predictions
   from \cite{Greene05} and from the current work.
   The dotted red line is the locus where the predictions
   are equal. 
   The black short-dash lines shows this
   $\pm 1\sigma$ envelope and the long-dash lines
   show the $\pm2.6\sigma$ envelope.
   {\it Note that this figure plots mass
   rather than virial product.}
   }
    \label{figure:GH05}
    \end{figure}

Here we compare our H$\alpha$-based
mass predictions with those previously published.
We consider first the early H$\alpha$-based
mass predictor of \cite{Greene05}; we rewrite their
equation (6) as
\begin{eqnarray}
\log M_{\rm GH05} & = & 7.331 + 
0.55\left[\log L({\rm H}\alpha) -42\right] \nonumber \\
& & +2.06\left[\log {\rm FWHM(H}\alpha) -3.5 \right].
\label{eq:GH05}
\end{eqnarray}
Figure \ref{figure:GH05} shows a direct comparison of
equations (\ref{eq:GH05}) and (\ref{eq:Hamass_FWHM})
for the sample in 
Tables \ref{table:RMDBHa} and \ref{table:SDSSHa}; note that in
Figure \ref{figure:GH05}, we are plotting mass 
rather than virial product. 
The best-fit results are given in 
line 1 of Table \ref{table:fits3}.
\cite{Greene05} rederived the relationship between the
H$\beta$ lag and the 5100\,\AA\ continuum, and
essentially reproduced the result of 
\cite{Kaspi05}. This was prior to the
first recognition that the contaminating starlight
needs to be accounted for prior to deriving this relationship \citep{Bentz06}; consequently the empirical relationship was steeper than the canonical value of 0.5. Empirical 
relationships among the 5100\,\AA\ continuum and
the H$\alpha$ and H$\beta$ emission line fluxes and
between the H$\alpha$ and H$\beta$ line widths
were also used. The scaling factor used
was $f = 3/4$ \citep{Netzer90}, which was what was widely used prior the first empirical calibration
\citep{Onken04}.

\begin{figure}
   \centering
   \resizebox{\hsize}{!}{\includegraphics{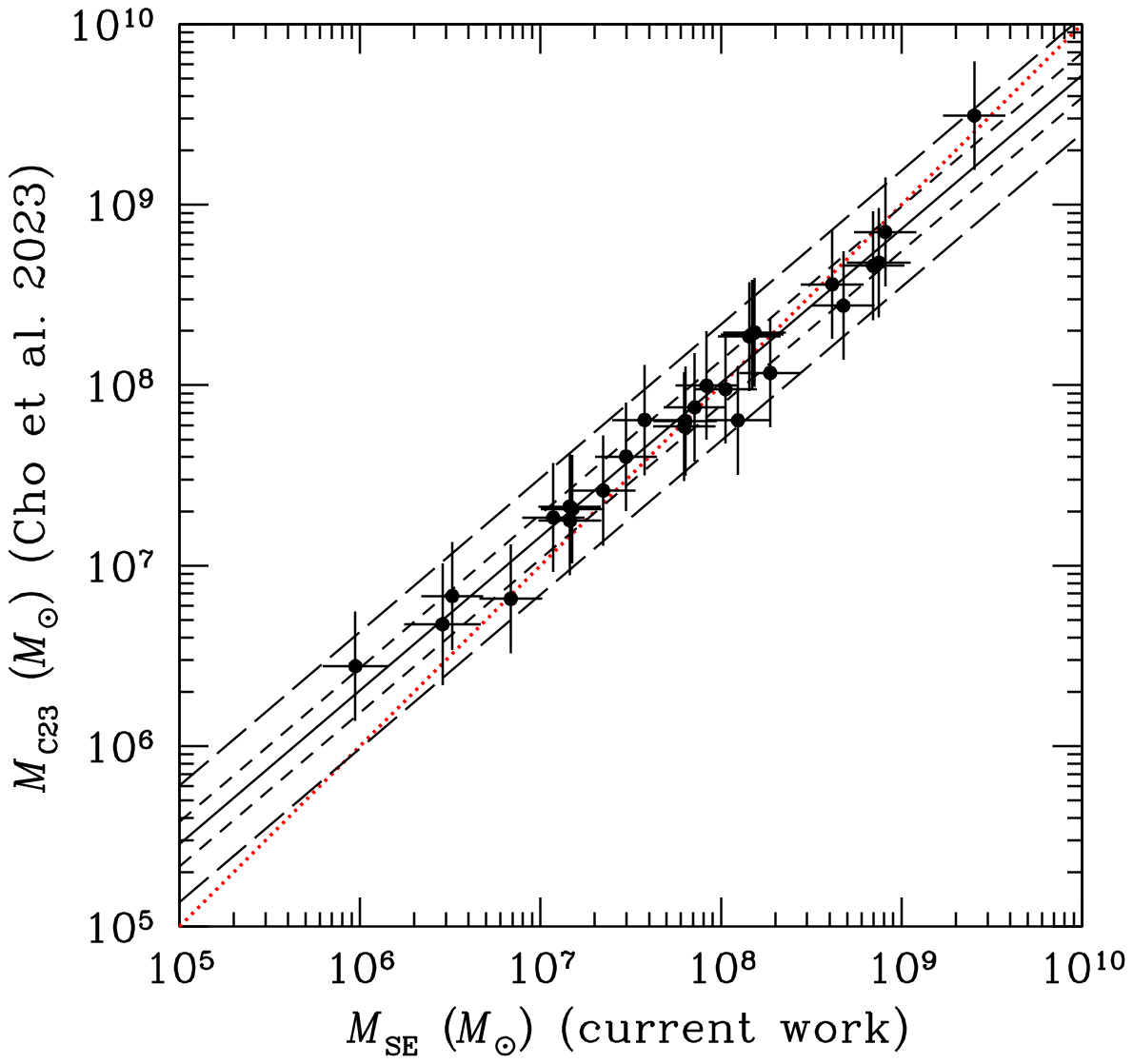}}
   \caption{Direct comparison of the mass predictions
   from \cite{Cho23} and from the current work.
   The dotted red line is the locus where the predictions
   are equal. 
   The black short-dash lines shows this
   $\pm 1\sigma$ envelope and the long-dash lines
   show the $\pm2.6\sigma$ envelope.
   {\it Note that this figure plots mass
   rather than virial product.}
   }
    \label{figure:Cho23}
    \end{figure}

We also consider comparison with a more recent effort
by \cite{Cho23}, whose database overlaps with ours considerably.
Their equation (6) can be written as
\begin{eqnarray}
\log M_{\rm C23} & = & 7.505 + 
0.61\left[\log L({\rm H}\alpha) -42\right] \nonumber \\
& & +2.0\left[\log {\rm FWHM(H}\alpha) -3.5 \right].
\label{eq:Cho23}
\end{eqnarray}
The predictions from this equation are compared with those of our equation (\ref{eq:Hamass_FWHM})
in Figure \ref{figure:Cho23}; note that we are plotting mass
rather than virial product.
The best-fit parameters are given in line 2
of Table \ref{table:fits3}.
Again, the slope of the relationship between
these two predictions is less than unity at least in part
because of the lack of an Eddington ratio
correction. Moreover, some of  
the underlying assumptions are so different. Salient points:
\begin{enumerate}
\item \cite{Cho23} assume a scaling factor value of 
$\log f = 0.05 \pm 0.12$ \citep{Woo15} when FWHM is used
as the line-width measure. This corrects 
${\rm FWHM}_{\rm M}$ to $\sigma_{\rm M}$ for the
{\it mean} ratio of 
$\langle {\rm FWHM}_{\rm M}/\sigma_{\rm M}\rangle$
\citep[cf.][]{Collin06}, but it does not account for the fact
that the relationship between ${\rm FWHM}_{\rm M}$ and 
$\sigma_{\rm M}$ is neither constant nor linear
\citep[e.g., Figure 9 of][]{Peterson14}.
Indeed, for the H$\alpha$ lines examined in this
investigation the width ratio cover the range
\begin{equation}
\nonumber 
0.78 \la {\rm FWHM}_{\rm M}/\sigma_{\rm M} \la 2.73,
\end{equation}
compared to the Gaussian value ${\rm FWHM}/\sigma = 2.35.$
\item The slope we find for the H$\alpha$ $R$--$L$ relationship,
$b = 0.497 \pm 0.016$, is shallower than their
slope, $b = 0.61 \pm 0.04$.
\item \cite{Cho23} assume that the mass scales as ${\rm FWHM}^2$
while we find that the dependence of mass on ${\rm FWHM}$ is
much shallower for H$\alpha$, as it is for H$\beta$
(Paper I).

\end{enumerate}   

  \begin{table*}
\caption{Fits to Comparisons: $y = a +b(x-x_0)$}
\label{table:fits3}
 \centering
 \begin{tabular}{cllcccccc}
 \hline\hline
 Line & $x$ & $y$ & $a \pm \Delta a$ & $b \pm \Delta b$ &
 $x_0$ & $\varepsilon_y$ & $\Delta$ & Figures \\
 (1) & (2) & (3) & (4) & (5) & (6) & (7) & (8) & (9) \\
 \hline
1 & $\log M_{\rm SE}$ (current)&
$\log M_{\rm SE}$ (GH05) &
$7.685 \pm 0.076$ &$0.802 \pm 0.011$ & 7.854 &
$<10^{-3}$ & 0.152 & \ref{figure:GH05} \\
2 & $\log M_{\rm SE}$ (current) 
& $\log M_{\rm SE}$ (Cho23) &
$7.890 \pm 0.064$ & $0.852 \pm 0.008$ & 7.754 &
$<10^{-3}$ & 0.124 & \ref{figure:Cho23} \\
  \hline
\end{tabular}
\end{table*}

\section{Conclusions}

In this contribution, we derive single-epoch
black hole mass estimators based on the luminosity
and line width of the broad H$\alpha$ emission line
(equations \ref{eq:Hamass_sigma} and 
\ref{eq:Hamass_FWHM}) with a typical
formal uncertain around 0.2 -- 0.4\,dex relative
to the reverberation masses, depending on which
emission line and line-width measure are used.
Both the H$\alpha$ and H$\beta$-based estimators
are calibrated over the luminosity range
$41 \la \log L_{\rm AGN}(5100\,{\rm \AA}) \la 46\,{\rm erg\,s}^{-2}$.
Our treatment takes into account the three
parameters known to affect black-hole mass:
luminosity, line width, and Eddington ratio.
As is the case with the H$\beta$ emission line
(Paper I), either line dispersion (equation
\ref{eq:sigmadef}) or FWHM can be used as the
line-width measure {\em though not interchangeably:}
the mass-dependence on line-width is shallower
for FWHM than for line dispersion.
    
\begin{acknowledgements}

Funding for the Sloan Digital Sky Survey IV has been provided by the
Alfred P.\ Sloan Foundation, the U.S.\ Department of Energy Office of
Science, and the Participating Institutions. SDSS-IV acknowledges
support and resources from the Center for High-Performance Computing
at the University of Utah. The SDSS web site is www.sdss.org. 
DG acknowledges the support from PROYECTOS FONDO de ASTRONOMIA ANID - ALMA 2021 Code :ASTRO21-0007.
EDB, SC, EMC, and AP acknowledge the support from MIUR grant PRIN 2017 20173ML3WW-001 and Padua University grants DOR 2021-2023; they are also funded by INAF through grant PRIN 2022 C53D23000850006.
WNB acknowledges support from NSF grant AST-2407089.
LCH was supported by the National Science Foundation of China (11721303, 11991052, 12011540375, 12233001), the National Key R\&D Program of China (2022YFF0503401), and the China Manned Space Project (CMS-CSST-2021-A04, CMS-CSST-2021-A06).
M.V. gratefully acknowledges financial support from the Independent Research Fund Denmark via grant numbers DFF 8021-00130 and  3103-00146.
\end{acknowledgements}

%
%

\begin{appendix}
    
\section{Data used in this analysis}

\begin{sidewaystable*}
     \caption{Reverberation Mapped AGNs (H$\alpha$)}
     \label{table:RMDBHa}
     \centering
     \begin{tabular}{lcccccccc}
       \hline\hline
     &  & & $\tau$(H$\alpha$) &
       $\log L({\rm H}\alpha)$ & ${\rm FWHM_{\rm M}(H}\alpha)$ &
       $\sigma_{\rm M}({\rm H}\alpha)$ &
       $\sigma_{\rm R}({\rm H}\alpha)$ &
       $\log \mu_{\rm RM}({\rm H}\beta)$\\
       Source & Referemces & $z$  & (days) & (erg\,s$^{-1}$) &
       (km\,s$^{-1}$) & (km\,s$^{-1}$) &(km\,s$^{-1}$) 
       & $(M_\odot)$\\
       (1) & (2) & (3) & (4) & (5) & (6) & (7) & (8) & (9) \\
       \hline
PG\,0026+129	& 1	& 0.14200	&$ 98.16^{+28.28}_{-25.48}	$&$ 43.524  \pm	0.041	$&$ 1532  \pm	32	$&$ 1769  \pm	17	$&$ 1961  \pm	135	$&$ 7.833 \pm	0.107$ \\
PG\,0052+251	& 1	& 0.15445	&$ 168.22^{+57.11}_{-48.74}	$&$ 43.738  \pm	0.046	$&$ 3605  \pm	137	$&$ 2054  \pm	12	$&$ 1913  \pm	85	$&$ 7.745 \pm	0.122$ \\
Mrk\,6		& 2,3	& 0.01881	&$ 28.42^{+7.09}_{-6.51}	$&$ 42.513  \pm	0.043	$&$ 5322  \pm	142	$&$ 2870  \pm	22	$&$ 2780  \pm	35	$&$ 7.522 \pm	0.078$ \\
Mrk\,6		& 2,3	& 0.01881	&$ 24.85^{+5.89}_{-7.06}	$&$ 42.548  \pm	0.033	$&$ 5322  \pm	142	$&$ 2870  \pm	22	$&$ 2780  \pm	35	$&$ 7.444 \pm	0.060$ \\
Mrk\,6		& 2,3	& 0.01881	&$ 34.88^{+12.50}_{-12.11}	$&$ 42.531  \pm	0.033	$&$ 5322  \pm	142	$&$ 2870  \pm	22	$&$ 2780  \pm	35	$&$ 7.440 \pm	0.137$ \\
PG\,0804+761	& 1	& 0.10000	&$ 170.93^{+14.26}_{-11.95}	$&$ 43.729  \pm	0.026	$&$ 2756  \pm	16	$&$ 2596  \pm	9	$&$ 2046  \pm	138	$&$ 8.047 \pm	0.061$ \\
Mrk\,110		& 4	& 0.03529	&$ 29.61^{+3.76}_{-5.18}	$&$ 42.486  \pm	0.043	$&$ 	  \ldots	$&$ 	  \ldots	$&$  	  \ldots	$&$ 6.770 \pm	0.128$ \\
Mrk\,142		& 5,6,7	& 0.04494	&$ 2.94^{+0.94}_{-1.11}		$&$ 42.085  \pm	0.034	$&$ 1350  \pm	39	$&$ 925	  \pm	28	$&$ 934	  \pm	61	$&$ 5.596 \pm	0.125$ \\
NGC\,3516		& 8	& 0.00884	&$ 13.35^{+7.08}_{-2.87}	$&$ \ldots  		$&$ 3238  \pm	27	$&$ 4152  \pm	185	$&$ 	  \ldots	$&$ 6.827 \pm	0.164$ \\
SBS\,1116+583A	& 5,6,7	& 0.02787	&$ 4.02^{+1.38}_{-0.95}		$&$ 41.251  \pm	0.030	$&$ 2059  \pm	359	$&$ 1250  \pm	39	$&$ 1218  \pm	125	$&$ 6.022 \pm	0.109$ \\
Arp\,151		& 5,6,7	& 0.02109	&$ 7.89^{+0.99}_{-0.86}		$&$ 41.525  \pm	0.053	$&$ :1852 \pm	7	$&$ 1367  \pm	11	$&$ 937	  \pm	34	$&$ 6.087 \pm	0.068$ \\
NGC\,3783		& 9,10	& 0.00973	&$ 13.32^{+3.56}_{-4.29}	$&$ 41.601  \pm	0.044	$&$ 	  \ldots	$&$ 	  \ldots	$&$ 	  \ldots	$&$ 6.787 \pm	0.128$ \\
Mrk\,1310		& 5,6,7	& 0.01956	&$ 4.52^{+0.67}_{-0.65}		$&$ 41.345  \pm	0.031	$&$ :561  \pm	136	$&$ 887	  \pm	80	$&$ 717	  \pm	75	$&$ 5.609 \pm	0.075$ \\
Mrk\,202		& 5,6,7	& 0.02102	&$ 22.49^{+1.80}_{-1.74}	$&$ 41.133  \pm	0.034	$&$ :463  \pm	38	$&$ 746	  \pm	77	$&$ 734	  \pm	22	$&$ 5.412 \pm	0.209$ \\
NGC\,4253		& 5,6,7	& 0.01293	&$ 24.87^{+0.85}_{-0.73}	$&$ 41.616  \pm	0.018	$&$ 1013  \pm	15	$&$ 801	  \pm	42	$&$ 726	  \pm	35	$&$ 6.408 \pm	0.054$ \\
PG\,1226+023	& 1	& 0.15834	&$ 256.83^{+40.88}_{-150.39}	$&$ 44.608  \pm	0.035	$&$ 3036  \pm	49	$&$ 2514  \pm	47	$&$ :2075 \pm	239	$&$ 8.277 \pm	0.117$ \\
PG\,1229+204	& 1	& 0.06301	&$ 65.13^{+55.20}_{-23.75}	$&$ 42.852  \pm	0.035	$&$ 2996  \pm	34	$&$ 1931  \pm	10	$&$ 1737  \pm	118	$&$ 7.149 \pm	0.257$ \\
NGC\,4748		& 5,6,7	& 0.01463	&$ 7.87^{+3.00}_{-4.64}		$&$ 41.681  \pm	0.031	$&$ :1967 		$&$ 901	  \pm	34	$&$ 1035  \pm	74	$&$ 5.670 \pm	0.153$ \\
PG\,1307+085	& 1	& 0.15500	&$ 162.43^{+67.49}_{-56.50}	$&$ 43.679  \pm	0.035	$&$ 3685  \pm	31	$&$ 2090  \pm	16	$&$ 1843  \pm	98	$&$ 7.834 \pm	0.176$ \\
NGC\,5273		& 11	& 0.00362	&$ 1.04^{+0.67}_{-0.80}		$&$ 40.318  \pm	0.071	$&$ 3032  \pm	54	$&$ 1781  \pm	36	$&$ 1783  \pm	66	$&$ 6.012 \pm	0.278$ \\
PG\,1411+442	& 1	& 0.08960	&$ 108.46^{+62.92}_{-47.40}	$&$ 43.410  \pm	0.015	$&$ 2247  \pm	44	$&$ 2675  \pm	13	$&$ 2437  \pm	196	$&$ 7.796 \pm	0.216$ \\
NGC\,5548		& 12,13	& 0.01718	&$ 17.17^{+2.54}_{-2.44}	$&$ 42.316  \pm	0.028	$&$ 	  \ldots	$&$ 	  \ldots	$&$ 1843  \pm	98	$&$ 7.038 \pm	0.049$ \\
NGC\,5548		& 5,6,7	& 0.01718	&$ 10.85^{+1.31}_{-1.15}	$&$ 42.060  \pm	0.028	$&$ :1643 \pm	12	$&$ 3540  \pm	40	$&$ 3562  \pm	141	$&$ 7.172 \pm	0.038$ \\
PG\,1426+015	& 1	& 0.08657	&$ 77.77^{+29.86}_{-30.76}	$&$ 43.407  \pm	0.031	$&$ 5534  \pm	73	$&$ 3563  \pm	26	$&$ 4254  \pm	290	$&$ 8.342 \pm	0.157$ \\
PG\,1613+658	& 1	& 0.12900	&$ 39.80^{+18.11}_{-14.82}	$&$ 43.625  \pm	0.026	$&$ 6297  \pm	85	$&$ 4136  \pm	65	$&$ 	  \ldots	$&$ 7.705 \pm	0.166$ \\
PG\,1617+175	& 1	& 0.11244	&$ 96.73^{+23.40}_{-26.91}	$&$ 43.224  \pm	0.034	$&$ 4503  \pm	134	$&$ 2780  \pm	10	$&$ 2483  \pm	160	$&$ 7.982 \pm	0.198$ \\
3C\,390.3		& 14	& 0.05610	&$ 51.65^{+1.85}_{-1.96}	$&$ 43.297  \pm	0.024	$&$ 12563 \pm	31	$&$ 4607  \pm	29	$&$ 4839  \pm	215	$&$ 8.431 \pm	0.041$ \\
NGC\,6814		& 5,6,7	& 0.00521	&$ 9.32^{+2.31}_{-1.50}		$&$ 41.015  \pm	0.028	$&$ 2909  \pm	3	$&$ 1686  \pm	29	$&$ 1082  \pm	52	$&$ 6.526 \pm	0.061$ \\
       \hline
     \end{tabular}
     \tablefoot{Column (1) is the common name of the source.
     Column (2) provides references for the data (below).
     Column (3) is the redshift. Column (4) is the H$\alpha$ time delay relative to the
     continuum, in the rest frame of the source.
     Column (5) is the luminosity of the H$\alpha$ emission line.
     Columns (6), (7), and (8) are
      the H$\alpha$ line width, characterized by
      FWHM in the mean spectrum, line dispersion in the
      mean spectrunm, and line dispersion in the RMS
      spectrum, respectively, all in the rest frame of the source.
      Column (9) is the H$\beta$ virial product from Paper I.
     }
     \tablebib{
       (1)~\citet{Kaspi00}; (2)~\citet{Sergeev99}; (3)~\citet{Doroshenko12}; (4) \citet{Kollatschny01}; (5) \citet{Bentz09b}, (6) \citet{Bentz10};
       (7) \citet{Walsh09}; (8) \citet{Wanders93}; (9) \citet{Reichert94}; (10) \citet{Stirpe94}; (11) \citet{Bentz14}; (12) \citet{Clavel91};
       (13) \citet{Dietrich93}; (14) \citet{Dietrich12}.
       }
     \end{sidewaystable*}
\begin{sidewaystable*}
     \caption{Reverberation Mapped AGNs (SDSS H$\alpha$)}
     \label{table:SDSSHa}
     \centering
     \begin{tabular}{lcccccccc}
       \hline\hline
       & & & $\tau$(H$\alpha$) &
       $\log L({\rm H}\alpha)$ & ${\rm FWHM_{\rm M}(H}\alpha)$ &
       $\sigma_{\rm M}({\rm H}\alpha)$ &
       $\sigma_{\rm R}({\rm H}\alpha)$ 
       & $\log \mu_{\rm RM}({\rm H}\beta)$\\
      RMID & $z$ & MJD Range & (days) & (erg\,s$^{-1}$) &
       (km\,s$^{-1}$) & (km\,s$^{-1}$) &(km\,s$^{-1}$) &$(M_\odot)$ \\
       (1) & (2) & (3) & (4) & (5) & (6) & (7) 
       & (8) & (9)  \\
       \hline
88	& 0.5175 &56358--57195	&$ 56.75^{+3.14}_{-4.35} $& $ 42.686\pm 0.026 $& $ 4899	\pm 1017$& $ 2048\pm 431 $ &$2596\pm 11 $&   $\ldots$ \\
160	& 0.3598 &56660--56838	&$ 18.01^{+5.51}_{-5.04} $& $ 42.467\pm 0.018 $& $ 4014	\pm 101 $& $ 1609\pm 15  $ &$1318\pm 6  $& $7.192 \pm	0.068$\\
191	& 0.4419 &56660--56838	&$ 17.39^{+2.37}_{-2.44} $& $ 42.005\pm 0.034 $& $ 2349	\pm 61	$& $ 907 \pm 21	 $ &$750 \pm 15 $& $6.245 \pm	0.105$\\
229	& 0.4698 &56660--56838	&$ 24.47^{+10.90}_{-5.10}$& $ 42.159\pm 0.017 $& $ 2555 \pm 196 $& $ 1437\pm 32  $ &$1702\pm 16 $& $7.001 \pm	0.103$\\
252	& 0.2809 &56660--56838	&$ 14.81^{+3.61}_{-3.05} $& $ 41.952\pm 0.014 $& $ 8016	\pm 184	$& $ 3079\pm 56  $ &$2880\pm 8  $&   $\ldots $ \\
272	& 0.2629 &56358--57195	&$ 28.95^{+9.39}_{-7.05} $& $ 42.665\pm 0.015 $& $ 2692	\pm 54	$& $ 1321\pm 13  $ &$1265\pm 5  $& $6.928 \pm	0.114$\\
320	& 0.2651 &56660--56838	&$ 20.58^{+5.13}_{-3.60} $& $ 42.068\pm 0.021 $& $ 3383	\pm 49	$& $ 1467\pm 36  $ &$1211\pm 6  $& $7.026 \pm	0.090$\\
371	& 0.4726 &56660--57195	&$ 14.56^{+3.68}_{-4.80} $& $ 42.607\pm 0.038 $& $ 2901	\pm 113	$& $ 1358\pm 49  $ &$1341\pm 8  $& $6.722 \pm	0.039$\\
377	& 0.3372 &56660--56838	&$  5.98^{+3.00}_{-4.13} $& $ 41.797\pm 0.027 $& $ 2970	\pm 67	$& $ 1257\pm 20  $ &$1184\pm 10 $& $6.562 \pm	0.039$\\
645	& 0.4739 &56660--57195	&$ 21.03^{+9.56}_{-10.90}$& $ 42.573\pm 0.024 $& $ 3245	\pm 98	$& $ 1377\pm 40  $ &$918 \pm 4  $& $6.921 \pm	0.048$\\
733	& 0.4547 &56660--56838	&$ 37.09^{+23.75}_{-9.95}$& $ 42.459\pm 0.029 $& $ 3797 \pm 95	$& $ 1548\pm 55  $ &$1517\pm 11 $& $\ldots$\\
768	& 0.2587 &56660--56838	&$ 18.34^{+3.09}_{-2.82} $& $ 42.256\pm 0.026 $& $ 6570	\pm 321	$& $ 2622\pm 63  $ &$2228\pm 4  $& $\ldots$\\
772	& 0.2491 &56660--57195	&$  5.83^{+3.62}_{-3.78} $& $ 41.880\pm 0.045 $& $ 2763	\pm 32	$& $ 1069\pm 15	 $ &$859 \pm 2  $& $5.903 \pm 0.101 $\\
776	& 0.1164 &56660--57195	&$  6.28^{+1.75}_{-2.37} $& $ 41.665\pm 0.028 $& $ 3038	\pm 49	$& $ 1418\pm 16  $ &$1097\pm 2  $& $6.609 \pm	0.071$ \\
840	& 0.2439 &56660--56838	&$  8.84^{+3.95}_{-1.95} $& $ 41.869\pm 0.052 $& $ 4908	\pm 283	$& $ 1953\pm 39  $ &$1703\pm 4  $& $7.287 \pm	0.127$ \\
      \hline
            \end{tabular}
      \tablefoot{Column (1) is RMID: correspondence with
      SDSS identifier can be found in Table 1 of
      \cite{Grier17b}. Column (2) is redshift. Column (3) is range of dates (MJD) for which the time-series analysis was undertaken.
      Column (4) is the H$\alpha$ lag relative to the continuum.
      Column (5) is the luminosity of the H$\alpha$ line.
      Columns (6), (7), and (8) are
      the H$\alpha$ line width, characterized by
      FWHM in the mean spectrum, line dispersion in the
      mean spectrunm, and line dispersion in the RMS
      spectrum, respectively, all in the rest frame of the source.
      Column (9) is the reverberation virial product
      for H$\beta$ from Paper I.
         }
\tablebib{Line widths and luminosities are from \citet{Wang19}.}
     \end{sidewaystable*}

\begin{sidewaystable*}
    \caption{Multivariate Relations: $z = a +b(x-x_0) + c(y-y_0)$}
    \label{table:multivariate}
    \centering
    \begin{tabular}{cllcccccccc}
    \hline\hline
    Line & $x$ & $y$ & $z$ & $a \pm \Delta a$ & $b \pm \Delta b$ &
$c \pm \Delta c$ & $x_0$ & $y_0$ & $\varepsilon_z$ & $\Delta$ \\
 \hline
1 & $\log L({\rm H}\alpha)$ & $ \log \sigma_{\mathrm M}({\rm H}\alpha)$ & $\log \mu_{\rm RM} ({\rm H}\beta)$ &
 $6.996 \pm 0.044$ & $0.501 \pm 0.064$ & $2.397 \pm 0.263$ &
 42.467 & 3.227 & $0.211 \pm 0.042 $ & 0.230\\
 2 & $\log L({\rm H}\alpha)$ & $ \log {\rm FWHM}_{\mathrm M}({\rm H}\alpha)$ & $\log \mu_{\rm RM}({\rm H}\beta)$ &
 $7.082 \pm 0.066$ & $0.583 \pm 0.085$ & $1.173 \pm 0.247$ &
 42.531 & 3.314 & $0.310 \pm 0.066$ & 0.322 \\
 \hline
 \end{tabular}
 \end{sidewaystable*}

\end{appendix}

\end{document}